\documentclass{article}

\usepackage{graphicx}
\usepackage{epsfig}
\usepackage{amsfonts}

 \usepackage{amssymb}
\usepackage{amsmath}
\usepackage{hyperref}
\usepackage{graphicx}       % Include figure files
\usepackage{dcolumn}        % Align table columns on decimal point
\usepackage{bm}             % bold math
\usepackage{amssymb}
\usepackage{amstext}
\usepackage{tensor}

\date{\today}

\newcommand{\insertplot}[5]{\begin{figure}
 \hfill\hbox to 0.05in{\vbox to #5in{\vfill
 \inputplot{#1}{#4}{#5}}\hfill}
 \hfill\vspace{-.1in}
 \caption{#2}\label{#3}
 \end{figure}}
 \newcommand{\inputplot}[3]{
 \special{ps: plotfile #1}
}
\newcounter{fig}

\newcommand{\ee}{\end{equation}}
\newcommand{\eea}{\end{eqnarray}}

\def\theequation{\arabic{equation}}

\def\theequation{\thesection.\arabic{equation}}

\newcommand{\gam}{\gamma}
\newcommand{\bet}{\beta}
\newcommand{\alp}{\alpha}

\newcommand{\sig}{\sigma}

\newcommand{\beq}{\begin{equation}}
\newcommand{\eeq}{\end{equation}}

\begin{document}

\title{\Large{\bf Einstein-Dirac-Maxwell  wormholes: 
\\
ansatz, construction and properties of symmetric solutions}}
 \vspace{1.5truecm}

\author{
{\large }%$^{\ddagger}$
{\ Jose Luis Bl\'azquez-Salcedo}$^{1}$,
{\ Christian Knoll}$^{2}$, 
and
{ E. Radu}$^{3}$
\\
\\
$^{1}${\small  Departamento de F\'isica Te\'orica and IPARCOS, Universidad Complutense de Madrid, E-28040 Madrid, Spain}
\\
$^{2}${\small  Institut f\"ur  Physik, Universit\"at Oldenburg, Postfach 2503,
	D-26111 Oldenburg, Germany,
}
\\ 
$^{3}${\small
Departamento de Matem\'atica da Universidade de Aveiro and}
\\
{\small  Centre for Research and Development  in Mathematics and Applications (CIDMA),} 
\\ 
{\small    Campus de Santiago, 3810-183 Aveiro, Portugal}
}

\maketitle

%\date{\today}
%\pacs{04.70.-s, 04.70.Bw, 04.50.-h}
\begin{abstract}
We present a  discussion of the 
traversable wormholes in Einstein-Dirac-Maxwell theory
recently reported in e-Print: 2010.07317. 
This includes a 
detailed description of the ansatz and junction condition,
together with an investigation of the
domain of existence of the solutions.
{In this study, we assume symmetry under interchange of the two asymptotically flat regions of a wormhole.} 
Possible issues and  limitations of the approach are also discussed. 
\end{abstract}

%\tableofcontents

%%%%%%%%%%%%%%%%%%%%%%%%%%%%%%%%%%%%%%%%%%%%%%%%%%
\section{Introduction}
%%%%%%%%%%%%%%%%%%%%%%%%%%%%%%%%%%%%%%%%%%%%%%%%%%

The attempts to construct particle-like fermionic solutions
have
 started with the work of  
Ivanenko \cite{Ivanenko},
Weyl \cite{Weyl},
Heisenberg
\cite{Heisenberg}
and Finkelstein et. $al.$
\cite{Finkelstein:1951zz,Finkelstein2},
which considered 
a Dirac field model with a %positive 
quartic self-interaction term.
A rigorous
numerical study of such solutions
has been done by Soler in Ref.
 \cite{Soler:1970xp}
(see also  Ref. \cite{Cazenave:1986pj} for a  proof of existence). 
The study of such localized configurations
was mainly motivated as an attempt to construct models of extended particles.

In some sense, this type of solutions is the Dirac counterpart
of the Q-balls  \cite{Coleman:1985ki} in a model with a complex self-interacting scalar field, 
sharing with them a variety of features 
\cite{Herdeiro:2020jzx}.
Moreover, this analogy goes even further. 
As proven by 
Finster, Smoller and  Yau 
\cite{Finster:1998ws}
the inclusion of self-gravity effects
leads to the existence of
particle-like solitonic solutions of Einstein-Dirac equations
even in the absence of a self-interaction term for the Dirac field 
(see also 
\cite{Brill:1957fx},
 \cite{Lee:1986tr}
for early  work in this direction).  
These Dirac solitons possess all basic properties of the mini-boson stars
\cite{Kaup:1968zz,Ruffini:1969qy},
in particular some of the configurations being stable.
Also, as with boson stars \cite{Pena:1997cy}, 
no Black Hole (BH) horizon can be added at the center of a (spherically symmetric) Dirac soliton
\cite{Finster:1998ju},
\cite{Finster:1999tt}.
Subsequent work
includes extensions of the model in 
\cite{Finster:1998ws} 
with U(1) \cite{Finster:1998ux}
or SU(2) 
\cite{Finster:2000ps}
gauged fermions, or
the study of Einstein-Dirac 
spinning configurations \cite{Herdeiro:2019mbz}.

Note that in all aforementioned studies,  the Dirac field was treated 
as  a {\it quantum wave function}, 
its fermionic nature being imposed at the level of the occupation number: at most a single
 particle, in accordance to Pauli's exclusion principle.
Thus the second  quantization effects are ignored,
while the gravitational field is treated purely classically.

The Finster-Smoller-Yau solutions
(together with their various generalizations)
 are topologically trivial,
with a spacetime geometry which is a deformed Minkowski one.
However, as found recently in Ref. 
\cite{Blazquez-Salcedo:2020czn},
a Dirac field allows for another class of solutions
which are absent in the usual models with bosonic fields -- the traversable  wormholes (WHs).
In some sense, these solutions provide 
an explicit realization of Wheeler's idea of
``{\it electric charge without charge}'' \cite{Wheeler}, possessing a variety
of interesting properties.

The subject of traversable WHs has entered General Relativity (GR)
 with
 the work  of Ellis \cite{Ellis:1973yv} 
and Bronnikov \cite{Bronnikov:1973fh}, 
enjoying
% constant attention ever since.
increasing interest over the last decades.
A characteristic feature of a traversable  WH
is that it
necessarily requires a matter content violating the null energy condition 
\cite{Morris:1988cz},
\cite{Visser},
\cite{Lobo:2016zle}
Then, 
restricting to  a field theory source 
and
a classical setting, the bosonic matter fields necessarily possess a non-standard Lagrangian
($e.g.$ 'phantom' fields 
\cite{Ellis:1973yv},
\cite{Bronnikov:1973fh},
\cite{phantom}).
Another possibility is 
 to consider extensions of gravity beyond GR
(see $e.g.$
\cite{Kanti:2011jz},
\cite{Barcelo:2000zf}).

The  novelty of Ref. \cite{Blazquez-Salcedo:2020czn}
was to show that the situation may change  for fermions, 
with the existence of traversable WH solutions of the Einstein-Dirac(-Maxwell) equations.
An $exact$ WH solution with ungauged,  massless fermions was also reported there,
although with
a spinor wave function which is not normalizable.
The main purpose of this work is to provide a detailed description of the $numerical$ solutions in 
 Ref. \cite{Blazquez-Salcedo:2020czn} (which possess finite mass, charge and a 
normalizable spinor wave function),
 with emphasis
on a number of technical details.
The paper is organized as follows.   
The Sections 2 and 3 deal  with the general framework of the solutions.
In particular, we discuss the issue of a symmetric Ansatz together with the  
junction condition at the WH throat.
The numerical results are presented in Section 4.
We end with Section 5, where the emerging picture is summarized.
 The Appendices contain details on the formalism used in the description of 
  fermions in a curved geometry, together with a description of
	the numerical approach.
An exact solution with ungauged, massless  spinors 
is also discussed there. 
 
%%%%%%%%%%%%%%%%%%%%%%%%%%%%%%%%%%%%%%%%%%%%%%%%%%%%%%%%%%%%%%%%%%%%%%%%%%%%%%
\section{The Einstein-Dirac-Maxwell action and field equations}
%%%%%%%%%%%%%%%%%%%%%%%%%%%%%%%%%%%%%%%%%%%%%%%%%%%%%%%%%%%%%%%%%%%%%%%%%%%%%%

We consider Einstein’s gravity minimally coupled
with two U(1)-gauged {relativistic} fermions  with equal mass,  
the spin of which is taken 
 to be opposite in order to satisfy spherical symmetry.
 Working in units with 
 $G=c=\hbar=1$,
 the action of the corresponding Einstein-Dirac-Maxwell (EDM) model reads 
 \begin{eqnarray}
 \label{action}
 S = \frac{1}{4\pi}
 \int \mathrm d^4 x \sqrt{-g} \, 
 \left[ \frac{1}{4}R+ \mathcal L_D-\frac{1}{4}F^2    \right] \, ,
 \end{eqnarray}
 where $R$ is the Ricci scalar of the metric 
 $g_{\mu\nu}$, 
 $F_{\mu\nu}=\partial_{\mu}A_\nu-\partial_{\nu}A_\mu$
 is the field strength tensor of the U(1) field $A_\mu$,
 and
 \begin{eqnarray}
 \label{Lagrangian_Dirac}
 \nonumber
 \mathcal L_D = 
 \sum\limits_{\boldsymbol{\epsilon}=1,2} 
 \left[ \frac{\mathrm i}{2} \overline{\Psi}^{[\boldsymbol{\epsilon}]} 
 \gamma^\nu \hat{D}_\nu \Psi^{[\boldsymbol{\epsilon}]} 
 - \frac{\mathrm i}{2} 
 \hat{D}_\nu \overline{\Psi}_{\boldsymbol{\epsilon}}  \gamma^\nu \Psi^{[\boldsymbol{\epsilon}]} 
 - \mu 
 \overline{\Psi}_{\boldsymbol{\epsilon}}  \Psi^{[\boldsymbol{\epsilon}]}  
 \right] , 
 \end{eqnarray}
 where 
 $\mu$  is the mass  of {\it both} spinors, 
$q$ is the gauge coupling constant 
and
$\hat{D}_\mu  =  \partial_{ \mu} - 
 \Gamma_\mu -
	\mathrm i q A_\mu .
	$
Also,
$\gamma^\nu$
 are the curved
 space gamma matrices, while $\Gamma_\mu$ are the spinor connection matrices.
Their expression is  given in  Appendix \ref{Dfld_frm},
 together with some other details on the spinor formalism, where we follow
 the notation and conventions in Ref. \cite{Dolan:2015eua},

 The resulting field equations are  
 \begin{eqnarray}
 \label{Einstein}
 &&
 R_{\mu \nu} -\frac{1}{2}R g_{\mu \nu}= 2T _{\mu \nu} ~{\rm with}~T _{\mu \nu}=T^{(D)}_{\mu \nu}+ T^{(M)}_{\mu \nu},~~
 \\
 \label{matter}
 &&
 (\gamma^\nu \hat{D}_\nu-\mu) \Psi^{[\boldsymbol{\epsilon}]} =0,~~\nabla_{\mu}F^{\mu \nu}=q  j^\nu,
 \end{eqnarray}
 with
 the current
 \begin{equation}
 j^{\nu}= j^{\nu [\boldsymbol{1}]}+ j^{\nu [\boldsymbol{2}]}, ~~{\rm where}~~
%
%\sum\limits_{\epsilon=1,2}
j^{\nu [\boldsymbol{\epsilon}]}=
\overline{\Psi}^{[\boldsymbol{\epsilon}]}\gamma^{\nu} \Psi^{[\boldsymbol{\epsilon}]}~,
 \end{equation}
 and the stress-energy tensor
\begin{equation}
 T^{(D)}_{\mu \nu} = 
\sum\limits_{\boldsymbol{\epsilon}=1,2}T^{(D){[\boldsymbol{\epsilon}]}}_{\mu \nu} 
= -\frac{\mathrm i}{4}\sum\limits_{\boldsymbol{\epsilon}=1,2} \left[ \overline{\Psi}^{[\boldsymbol{\epsilon}]} \gamma_{\mu} 
\hat{D}_{\nu} \Psi^{[\boldsymbol{\epsilon}]} 
+ \overline{\Psi}^{[\boldsymbol{\epsilon}]} \gamma_{\nu} \hat{D}_{\mu} \Psi^{[\boldsymbol{\epsilon}]} 
- \hat{D}_{\nu}\overline{\Psi}^{[\boldsymbol{\epsilon}]} \gamma_{\mu} \Psi^{[\boldsymbol{\epsilon}]} 
- \hat{D}_{\mu}\overline{\Psi}^{[\boldsymbol{\epsilon}]} \gamma_{\nu} \Psi^{[\boldsymbol{\epsilon}]} \right],
 \end{equation}
\begin{equation}
 T^{(M)}_{\mu\nu} =  F_{\mu\alpha} F_\nu^{\alpha} - \frac{1}{4} F^2 g_{\mu\nu}.
\end{equation} %

%%%%%%%%%%%%%%%%%%%%%%%%%%%%%%%%%%%%%%%%%%%%%%%%%%%%%%%%%%%%
\section{Spherically symmetric wormholes: the framework}
%%%%%%%%%%%%%%%%%%%%%%%%%%%%%%%%%%%%%%%%%%%%%%%%%%%%%%%%%%%%

%%%%%%%%%%%%%%%%%%%%%%%%%%%%%%%%%%%%%%%%%%%%%%%%%%%%%%%%%%%%
\subsection{The metric}
%%%%%%%%%%%%%%%%%%%%%%%%%%%%%%%%%%%%%%%%%%%%%%%%%%%%%%%%%%%%

 Restricting to static, spherically-symmetric
solutions of the field equations, we consider a general metric ansatz
\begin{equation}
\label{metric}
ds^2=g_{tt}(r)dt^2+g_{rr}(r)dr^2+g_{\Omega \Omega}(r)d\Omega^2 = -F_0^2(r)dt^2+F_1^2(r)^2dr^2+F_2^2(r)d\Omega^2,
\end{equation}
where 
$d\Omega^2=d\theta^2+\sin^2\theta d\varphi^2$
is the line element on the two-sphere ($\theta$ and $\varphi$
being spherical coordinates with the usual range), 
while $r$ and $t$ are the radial and time coordinates, respectively.

Particle-like  (topologically trivial) solitons in 
EDM theory, 
are usually studied in Schwarzschild-like coordinates,  
with $0\leq r<\infty$ 
and $F_2(r)=r$.
However, the situation is more complicated for a WH geometry.
The characteristic feature here is the existence of two asymptotically flat regions
(the two sphere
$r=const.$, $t=const.$
possessing a minimal, nonzero size),
together with the absence of an event horizon.

A metric gauge choice which 
makes transparent the WH structure is
\begin{eqnarray}
\label{gc}
F_2(r)=\sqrt{r^2+r_0^2},~~~{\rm with}~~-\infty<r<\infty,
\end{eqnarray} 
where $r_0>0$ an input parameter--the radius of the throat.
% (which is located at $r=0$).

Given the above ansatz,
we choose the following 
  vierbein  (with $F_i>0$):
\begin{eqnarray}
e^r= \epsilon_r F_1 dr,~~e^\theta= F_2 d\theta,~~e^\varphi= F_2 \sin \theta d\varphi,~~
e^t= \epsilon_t F_0 dt,
\end{eqnarray}
where
\begin{eqnarray}
\epsilon_r =\pm 1,~~\epsilon_t =\pm 1.
\end{eqnarray}
The natural choice for particle-like configurations
\cite{Finster:1998ws},
\cite{Finster:1998ux},
is $\epsilon_r=1$.
However, for a WH geometry, 
one takes instead
\begin{eqnarray}
\label{eps}
\epsilon_r=1~~{\rm for}~~r>0~~~{\rm and}~~~\epsilon_r=-1~~{\rm for}~~r<0~,
\end{eqnarray}	
a choice which takes into account
the
 sign change of $r$ at the WH’s throat 
\cite{Cariglia:2018rhw}.
 
%%%%%%%%%%%%%%%%%%%%%%%%%%%%%%%%%%%%%%%%%%%%%%%%%%%%%%%%%%%%
\subsection{The matter functions}
%%%%%%%%%%%%%%%%%%%%%%%%%%%%%%%%%%%%%%%%%%%%%%%%%%%%%%%%%%%%
In our work, we 
 consider  a purely electric
Maxwell field with 
\begin{eqnarray}
A=V(r) dt,
\end{eqnarray}
 $V(r)$ being the electrostatic potential.

The Ansatz for the Dirac fields is 
chosen such that $i)$ it allows for spherically symmetric geometries,
and ii) it leads to a stress tensor and field equations which are compatible with
the 'reflection' symmetry $r\to -r$.
A Dirac ansatz which satisfies these 
conditions can be written in terms
 of a single complex function $z$, with
({see also Appendix \ref{Deq_sphr}})
 \begin{eqnarray}
 \label{Dirac-p}
&&
\Psi^{[1]} =
\begin{pmatrix} 
z(r) \cos(\frac{\theta}{2}) 
\\ 
\mathrm i \bar z(r)  \sin(\frac{\theta}{2}) \kappa 
\\ 
- \mathrm i\bar z(r)  \cos(\frac{\theta}{2}) 
\\ 
- z(r) \sin(\frac{\theta}{2})  \kappa
\end{pmatrix}
e^{\mathrm i(\frac{1}{2}\varphi-w t) } \ , \qquad
\Psi^{[2]} 
%\equiv  
%\Psi_+^{[2]} 
= \begin{pmatrix} 
\mathrm i z(r) \sin(\frac{\theta}{2})
\\ 
\bar z(r) \cos(\frac{\theta}{2})  \kappa 
\\ 
 \bar z(r)  \sin(\frac{\theta}{2})
\\ 
\mathrm i   z(r)  \cos(\frac{\theta}{2})  \kappa
\end{pmatrix}
e^{\mathrm i(-\frac{1}{2}\varphi-w t) } \ ,
\end{eqnarray}
 where
$\kappa=\pm 1$ and
$w$ is the field frequency.
Also,
 we note 
\begin{eqnarray}
\label{z}
z(r)=P(r)+\mathrm i Q(r)~,
\end{eqnarray}
 $P(r)$, $Q(r)$ being two real functions
subject to some conditions discussed below.
 $z$ can also be exprssed in terms of an amplitude $|\phi_0|$
and a phase $\alpha$,
\begin{eqnarray}
\label{za}
z(r)=|\phi_0|e^{\mathrm i\alpha}~,~~{\rm with}~~|\phi_0|=\sqrt{P^2+Q^2},~~\tan \alpha=\frac{Q}{P}~.
\end{eqnarray}

%%%%%%%%%%%%%%%%%%%%%%%%%%%%%%%%%%%%%%%%%%%%%%%%%%%%%%%%%%%%
\subsection{The  equations}
%%%%%%%%%%%%%%%%%%%%%%%%%%%%%%%%%%%%%%%%%%%%%%%%%%%%%%%%%%%%

Given the above ansatz, 
the Einstein equations read 
(where the prime denotes the derivative with respect to $r$):
\begin{eqnarray}
\nonumber
&& 
\frac{F_0'' }{F_0}+\frac{F_0' }{F_0}
(
\frac{F_2' }{F_2}-\frac{F_1' }{F_1}
)
-\frac{F_2'^2 }{2F_2^2}
+\frac{F_1^2 }{2F_2^2}
%-8\pi G
-2
\left(
\frac{3V'^2}{4F_0^2}
+2F_1^2
\big(
\epsilon_t (w+q V)
\frac{(P^2+Q^2)}{F_0}
-\frac{\kappa (P^2-Q^2)}{F_2}
\big)
\right)=0~,
\\
&&
\label{eqE}
\frac{F_2'' }{F_2}
+\frac{F_2' }{F_2} ( \frac{F_2' }{2F_2} -\frac{F_1' }{F_1} )
-\frac{F_1^2 }{2F_2^2}
+
\frac{1}{F_0}
\left(
\frac{V'^2}{2F_0 }+ 4 \epsilon_t (w+qV) F_1^2 (P^2+Q^2)
\right)=0~,
\\
&&
\nonumber
\frac{2F_0' F_2'}{F_0F_2}+\frac{F_2'^2-F_1^2}{F_2^2}
+2
\left(
\frac{V'^2}{2F_0^2}+ 4 \epsilon_r F_1(PQ'-QP') 
\right)=0~.
\end{eqnarray}
The spinor functions $P$, $Q$ satisfy the first order equations, 
\begin{eqnarray}
\label{eqD}
&&
\epsilon_r P' 
 +\epsilon_r ( \frac{F_0'}{2F_0}+ \frac{F_2'}{F_2})P
+\frac{F_1}{F_2} \left( \kappa-\frac{F_2}{F_0} \epsilon_t (w+q V) \right)Q -\mu F_1 P=0~,
\\
&&
\nonumber
\epsilon_r Q' 
+\epsilon_r ( \frac{F_0'}{2F_0}+ \frac{F_2'}{F_2})Q
+\frac{F_1}{F_2} \left( \kappa + \frac{F_2}{F_0} \epsilon_t (w+q V) \right)P +\mu F_1 Q=0~.
\end{eqnarray}
Finally, the Maxwell  equations reduce to a second order equation for the 
electrostatic potential
\begin{eqnarray}
\label{eqM}
 (\frac{F_2^2 V'}{F_0 F_1})'=4  \epsilon_t q F_1 F_2^2  (P^2+Q^2) .
\end{eqnarray}
Note that above equations are left invariant
by 
the transformation 
\begin{eqnarray}
\label{gT}
w\to w-\beta,~~
V \to V+\beta/q~,
\end{eqnarray}
with $\beta$ an arbitrary constant.

%%%%%%%%%%%%%%%%%%%%%%%%%%%%%%%%%%%%%%%%%%%%%%%%%%%%%%%%%%%%%%%%%
\subsection{The  'reflection' symmetry 
  and the junction condition}
%%%%%%%%%%%%%%%%%%%%%%%%%%%%%%%%%%%%%%%%%%%%%%%%%%%%%%%%%%%%%%%%%

The WH consists in two different regions 
$\Sigma_\pm$.
The `up' region ($\Sigma_+$) is found for $0<r<\infty$,
while the 'down' region ($\Sigma_-$) has $-\infty<r<0$.
These regions are joined at $r=0$, which is the position of the throat.

In this work, we are   interested in geometries
which are invariant under a reflexion with respect 
to the throat, $r \to -r$. 
Therefore
the metric functions and the energy-momentum tensor satisfy the conditions
\begin{eqnarray}
\label{cond-g}
F_i(-r) = F_i(r),~~(i=0,1,2)~~{\rm and}~~T_{\mu}^\nu (r)=T_{\mu}^\nu (-r).
\end{eqnarray} 
As for the spinor functions, we impose the following condition\footnote{Note that this is not the only possibility.
Similar results are found when considering instead a different Dirac-ansatz 
for the 'down'-region (still in terms of two real functions $(P,Q)$), 
together with a different identification instead of (\ref{transf}) 
\cite{Blazquez-Salcedo:2020czn}.
However, the picture becomes more complicated in this case, 
with a discontinuity of the spinors' phase
at the throat.
However, this is not  an issue \cite{Cariglia:2018rhw},
since 
at the junction, the 'up and 'down'
spinors are still related via an unitary transformation.
}
\begin{eqnarray}
\label{transf}
P(-r)= P(r),~~Q(-r)=Q (r).
%~~{\rm with }~~r>0.
\end{eqnarray} 
With these assumptions,
it is straightforward to verify
that the equations (\ref{eqE})- (\ref{eqM})
 remain invariant under the 
transformation 
$r \to -r$,
taken together with  (\ref{eps}) and (\ref{transf}) 
Here, we also assume that the product
$\epsilon_t (w+qV)$
does not change sign.
With respect to this, one distinguishes
two possibilities.
The first one is to take
 \begin{eqnarray}
\label{p1}
i)~~~
 \epsilon_t =1~~{\rm and }~~w\to w, ~~V\to V~~{\rm as}~~~r\to -r.
\end{eqnarray}
The second choice is rather unusual, employing a {\it time reversed} frame in the $\Sigma_-$-region,
 $i.e.$ with $t \to -t$ and
 \begin{eqnarray}
\label{p2}
ii)~~~
 \epsilon_t \to -\epsilon_t~~{\rm and }~~w\to -w, ~~V\to -V~~{\rm as}~~~r\to -r,
\end{eqnarray}
(with the usual choice  $ \epsilon_t=1$ for $r>0$).
Note that the product $wt$ (which enter the spinor phase), as well as the one form $A=V(r)dt$
are invariant for both choices above.

However, the distinction between 
the
possibilities
$i)$
and
$ii)$
above
does not manifest 
at level of construction of solutions,
together with their basic properties.  
Also, let us remark that the identification (\ref{transf})
does not leads to a discontinuity of the amplitude or the phase of the spinor function $z$
at the throat, $r=0$. 

Turning now to the joining at $r=0$
of the line elements for $\Sigma_\pm$ regions,
one remarks that in general this
is not `smooth', 
with a discontinuity of the metric derivatives.
This implies
 the presence of a thin mass shell structure at the throat,
with   a $\delta$-source 
($i.e.$ a thin matter shell)
added to the action (\ref{action}).
To get insight into this aspect, we evaluate
the second fundamental form
\begin{eqnarray}
K_{\mu \nu}=\frac{1}{2} (\nabla_\mu n_\nu+\nabla_\nu n_\mu)
\end{eqnarray}
 at $r=0^\pm$,
with $n_\nu$ the unit vector normal
at the surface $r=0$.
A straightforward computation shows that, at the throat, 
the only nonvanishing component of $K_{\mu \nu}$
is 
\begin{eqnarray}
K_{tt} \big|_{r=0^\pm}=\frac{F_0 F_0'}{F_1}\bigg|_{r=0^\pm}~.
\end{eqnarray}
However,
for
all solutions reported in this work,
 the first derivative of the metric function $F_0$
vanishes
at $r=0$.
As such, $K_{\mu \nu}=0$
and no extra-matter distribution 
at the throat  is required from this direction.

%%%%%%%%%%%%%%%%%%%%%%%%%%%%%%%%%%%%%%%%%%%%%%%%%%%%%%%%%%%%
\subsection{Asymptotics and boundary conditions }
%%%%%%%%%%%%%%%%%%%%%%%%%%%%%%%%%%%%%%%%%%%%%%%%%%%%%%%%%%%%
For  $r\to \pm \infty$, 
the Minkowski spacetime geometry is approached,
the spinor functions vanish, while the electric potential
approached a constant value
 \begin{eqnarray}
\label{inf1}
F_0 \big|_{r=\pm \infty}=
F_1 \big|_{r=\pm \infty}=1,~~P \big|_{r=\pm \infty}=Q \big|_{r=\pm \infty}=0,~~V \big|_{r=\pm \infty}= \epsilon_t \Phi.
\end{eqnarray}
%with $\Phi$ the electrostatic potential.
At the throat, one imposes
 \begin{eqnarray}
\label{r01}
F_0 \big|_{r=0^\pm}=F_{00},~~F_1 \big|_{r=0^\pm}=F_{10},~~V \big|_{r=0^\pm}=0,
\end{eqnarray}
(note that the condition for a vanishing electric potential fixes 
the residual gauge freedom (\ref{gT})), 
while
 \begin{eqnarray}
\label{r02}
P \big|_{r=0^\pm}=p_0,~~Q \big|_{r=0^\pm}=q_0,~~
%P \big|_{r=0^-}=-q_0,~~Q \big|_{r=0^-}=p_0,~~
\end{eqnarray}
with 
$F_{00}>0$,
$F_{10}>0$,
and
$p_0$,
$q_0$
arbitrary 
constants.
To simplify the picture,
we have restricted our numerical study to solutions with 
$p_0=-q_0$.

The solutions interpolating between the above asymptotics are found numerically, as described below.
However, 
one can construct an approximate local solution compatible with the above condition.
For example, the first terms in a
large-$r$ expression are
\begin{eqnarray}
F_0 \to 1-\frac{M}{r}+\dots,~~
F_1 \to 1+\frac{M}{r}+\dots,~~
V \to \Phi-\frac{Q_e}{r},~~
 P \to \frac{p_\infty}{r} e^{-\mu_{*}r}+\dots ,~~ Q\to \frac{q_\infty}{r} e^{-\mu_{*}r}+\dots,~~
\end{eqnarray}
with $M$ and $Q_e$ the
mass and electric charge.
Also,
$p_\infty =  -c_\infty (\frac{\mu_*}{\mu-w_*}-1)$, 
$q_\infty = c_\infty (\frac{\mu_*}{\mu-w_*}+1)$  
(with $c_\infty$ a constant and $w_*=w+q \Phi $).
In the above relations we note
$\mu_*=\sqrt{\mu^2-w_*^2}$, 
with
the bound state condition $\mu^2 > w_*^2$.
 
A local solution can also be constructed close to the throat,
 as a power series in $r$.
For example, the   metric functions behave as\footnote{
The leading order terms which enter the near-throat expansion satisfy the constraints
(with $v_1=V'(0^+)$):
 \begin{eqnarray}
\label{er01-s}
\frac{1}{r_0^2}=8 q_0^2 F_{10}^2(\mu-\frac{2w}{F_{00}}) ,~~
8q_0^2\left (\mu+(1-F_{10}^2)(\mu-\frac{2w}{F_{00}}) \right)+\frac{v_1^2}{F_{00}^2 F_{10}^2}=0.
\end{eqnarray} 
}
 \begin{eqnarray}
\label{er01}
F_0(r)=F_{00}+F_{02}r^2+\dots,~~F_1(r)=F_{10}+ \epsilon_r F_{11} r+F_{12}r^2+\dots~, ~~ 
\end{eqnarray} 
where
 $F_{02}$, 
 $F_{12}$
 are complicated expressions
in terms  of the 
input parameters and the 
values  at $r=0$ of various functions,
while
$F_{11}$
is a undetermined constant. 
Note that a value $F_{11}\neq 0$
implies discontinuity 
at the throat
for the first derivative of
metric function $g_{rr}$.
A systematic investigation of 
  all
numerical
 solutions constructed so far reveals that all of them have $F_{11}\neq 0$.
Therefore this feature
 seems to be generic, provided that the solutions 
are required to be symmetric around the throat\footnote{Note that the constant
$F_{11}$ does not enter the expression of
the 
 Riemann tensor evaluated at the throat.
As for curvature invariants, one finds 
\begin{eqnarray}
R\big |_{r=0^\pm}= \frac{2}{r_0^2}
\left(1-\frac{2}{F_{10}^2}-\frac{2F_{02}r_0^2}{F_{00}F_{10}^2} \right),~~
K\big |_{r=0^\pm}=\frac{4}{F_{10}^4}
\left(1-\frac{2+F_{10}^2}{r_0^4}+\frac{4F_{02}^2}{F_{00}^2}\right),~~
\end{eqnarray}
for the Ricci and Kretschmann scalars, respectively.
Moreover,
when expressing the line-element in terms of 
the
normal coordinate  to the throat 
$\eta=  \int F_1 dr$
(with 
 $g_{\eta\eta}=1$ and 
the following expression for small $|r|$:
$\eta =F_{10} r+\epsilon_r F_{11}r^2/2+\dots$),
one finds the first derivatives of both  
$g_{tt}(\eta)$ and $g_{\Omega \Omega}(\eta)$
vanish at the throat.
}.
 On the other hand, solutions with $F_1'(0)=0$ and $F_0'(0)=0$
appear to exist in a model with asymmetric wormholes \cite{Konoplya:2021hsm} (suggested also in \cite{Bronnikov:2021jlz}). 

\medskip

As for the electrostatic potential,
all solutions studied so far have 
a nonzero electric field at the throat,
$V'(0)\neq 0$.
For the (usual) choice (\ref{p1})
with $\epsilon_t=1$
and $V(r)=V(-r)$,
this implies 
that the electric field is discontinuous at $r=0$, with
 \begin{eqnarray}
V(r)= \epsilon_r v_1 r+\dots,~~{\rm and }~~F_{rt}\big|_{r=0^+}=-F_{rt}\big|_{r=0^-}=v_1.
\end{eqnarray}
The jump in the electric field  at the throat implies the presence
of a thin shell of electric charge located at $r=0$.
That is, for consistency, the action (\ref{action})
should be supplemented with a term
 \begin{eqnarray}
S_\Sigma= \frac{1}{4\pi}
 \int \mathrm d^4 x \sqrt{-g}   A_\nu J^\nu ,~~~{\rm with}~~J^\nu=\sigma_0  u^\nu \delta(r)
 \end{eqnarray} 
and the unit vector $u^\nu=\delta_t^\nu/F_0$.
 Also, $\sigma_0$ is the throat charge density,
 \begin{eqnarray}
\sigma_0= \frac{2v_1}{F_1^2F_0},
 \end{eqnarray} 
as resulting from a 
 straightforward computation.
The ($r=0$ localized) stress-energy tensor associated with this charge distribution is
 \begin{eqnarray}
T_{\alpha \beta}^{(s)}=A_{\alpha} \tilde J_\beta+A_{\beta} \tilde J_\alpha-h_{\alpha \beta } A_\nu \tilde J^\nu,
 \end{eqnarray}
where $h_{\alpha\beta}$ denotes the three-dimensional induced metric on the throat and 
$\tilde J^\nu =\sigma u^\nu$.
Since the U(1)-potential is vanishing at the throat,  $T_{\alpha \beta}^{(s)}$
does not contribute to the
 Einstein equations evaluated at $r=0$.
 
\medskip
 
A different picture is found for 
  the choice (\ref{p2}) of the mapping between $\Sigma_\pm$-regions,
with $V(-r)=-V(r)$ and the near-throat expression
$V(r)= v_1 r +O(r^2)$.
As a result, 
$
F_{rt}\big|_{r=0^+}=F_{rt}\big|_{r=0^-},
$
in which case no extra-matter 
exists at the WH throat.

%%%%%%%%%%%%%%%%%%%%%%%%%%%%%%%%%%%%%%%%%%%%%%%%%%%%%%%%%%%%
\subsection{Quantities of interest and a Smarr law}
%%%%%%%%%%%%%%%%%%%%%%%%%%%%%%%%%%%%%%%%%%%%%%%%%%%%%%%%%%%%

The only global charges 
of the solutions  
are the mass $M$ and the electric charge $Q_e$.
The WHs also possess a nonzero  throat area  
 \begin{eqnarray}
\label{th}
A_t=4\pi r_0^2~.
\end{eqnarray}
Also, for each spinor, one defines a Noether charge
$Q_N$. 
Restricting to the $\Sigma_+$-region, the expression of $Q_N$ reads
	\begin{eqnarray}
\label{QN} 
Q_{N}^{[\boldsymbol{1}]}= 
Q_{N}^{[\boldsymbol{2}]}= Q_{N}=
\frac{1}{4\pi}\int_{\Sigma_+} d^3x \sqrt{-g}j^{t [\boldsymbol{\epsilon}] } 
=2 \int_0^{\infty} dr F_1 F_2^2 (P^2+Q^2).
\end{eqnarray}
 By integrating the Maxwell equations,
one finds 
	\begin{eqnarray}
	\label{relQ}
	Q_e=2 q Q_N+Q_{T},
	\end{eqnarray}
with
$Q_T=V'(0^+) r_0^2/(F_0(0)F_1(0))$.

The WHs satisfy a Smarr law,
the mass being the sum of an electrostatic term and a
 bulk contribution 
\begin{eqnarray}
M= \Phi Q_e+M_{(B)},
\end{eqnarray}
with
\begin{eqnarray}
\nonumber
&
M_{(B)}=4\int_0^{\infty}
dr   F_1 F_2^2
\big[
2\mu F_0 P Q
+(2w+q V)(P^2+Q^2)
\big].
\end{eqnarray}
{
Similar relations hold for the $\Sigma_-$-region, in agreement with
the reflection symmetry of the solutions.
}

%%%%%%%%%%%%%%%%%%%%%%%%%%%%%%%%%%%%%%%%%%%%%%%%%%%%%%%%%%%%
\subsection{Scaling symmetry and one particle condition}
%%%%%%%%%%%%%%%%%%%%%%%%%%%%%%%%%%%%%%%%%%%%%%%%%%%%%%%%%%%%

The equations of the model are invariant under the scaling 
transformation
(the variables and quantities
which are not specified remain invariant):
	\begin{eqnarray}
	\label{one1}
 (r,r_0)\to \lambda (r,r_0), ~~~
(P,Q) \to (P,Q)/ \sqrt{\lambda},~~~
(\mu,q,w) \to (\mu,q,w)/\lambda,
\end{eqnarray}
where $\lambda$ is a positive constant, 
while various quantities of interest transform as
\begin{eqnarray}
\label{one}
(M,Q_e)\to \lambda (M,Q_e),~(Q_{N },A_t) \to \lambda^2 (Q_{N },A_t).
\end{eqnarray} 

As with the Einstein-Dirac(-Maxwell) solitons 
\cite{Finster:1998ws},
\cite{Finster:1998ux},
\cite{Herdeiro:2017fhv},	
this transformation is used to impose the one particle condition,
$Q_N=1$,
for {\it each}  spinor in 
  both 
'up' or 'down' regions.
That is, solving numerically the
field equations with
some input values of $\{ \mu,q,r_0,w \}$
one finds a solution with a nonzero $Q_N^{(num)}$.
Then the physical solution with 
$Q_N=1$
results from 
(\ref{one1}),
(\ref{one}),
with $\lambda=1/\sqrt{Q_N^{(num)}}$.

Let us also remark that only quantities which are invariant under the transformation 
(\ref{one1}),
(\ref{one})
	(like 
	$M/Q_e$
	or $A_t/Q_e^2$)
	are relevant.

%%%%%%%%%%%%%%%%%%%%%%%%%%%%%%%%%%%%%%%%%%%%%%%%%%%%%%%%%%%%%%%%%%%%%%%%%%%%%%
\section{Numerical solutions}
%%%%%%%%%%%%%%%%%%%%%%%%%%%%%%%%%%%%%%%%%%%%%%%%%%%%%%%%%%%%%%%%%%%%%%%%%%%%%%
%%%%%%%%%%%%%%%%%%%%%%%%%%%%%%%%%%%%%%%%%%%%%%%%%%%%%%%%%%%%%%%%%%%%%%%%%%%%%%

In this section we analyze the properties of the symmetric and smooth WHs that are obtained within the setting described in the previous sections.
In order to obtain the WH solutions, we solve numerically the field equations, imposing the boundary conditions that follow from the expansions at infinity and around the throat.
 More details on the specific parametrization, the equations that are solved
in practice and numerical solver are provided in Appendix \ref{det_num}. 
 
As mentioned above,
all solutions discussed here are symmetric with respect to a reflection at the throat, 
and  satisfy the condition $F_0'(0)=0$,
{while
 $F_1'(0^+)=-F_1'(0^-)\neq 0$. 
Also, to simplify the picture, 
we shall restrict our study to fundamental solutions 
($i.e.$
no radial excitations of the spinors)
and, moreover,  we shall display the profiles of various functions of interest
 for $r \geq 0$ only.
}

%%%%%%%%%%%%%%%%%%%%%%%%%%%%%%%%%%%%%%%%%%%%%%%%%%%%%%%%%%%%%%%%%%%%%%%%%%%%%%
\subsection{Solutions' properties}
%%%%%%%%%%%%%%%%%%%%%%%%%%%%%%%%%%%%%%%%%%%%%%%%%%%%%%%%%%%%%%%%%%%%%%%%%%%%%%

\begin{figure}
	\centering
	\includegraphics[height=3.0in,angle=-90]{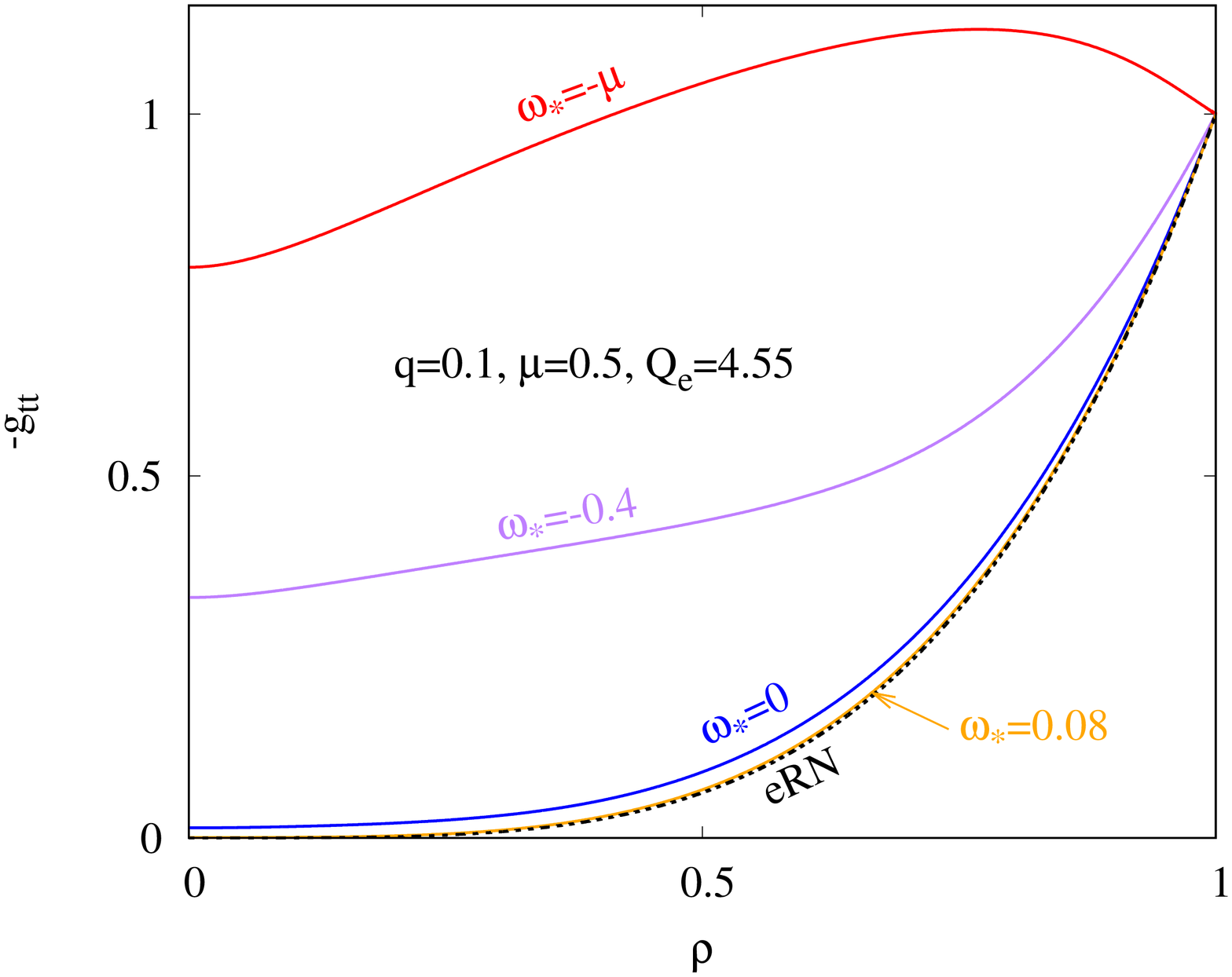} 
	\includegraphics[height=3.0in,angle=-90]{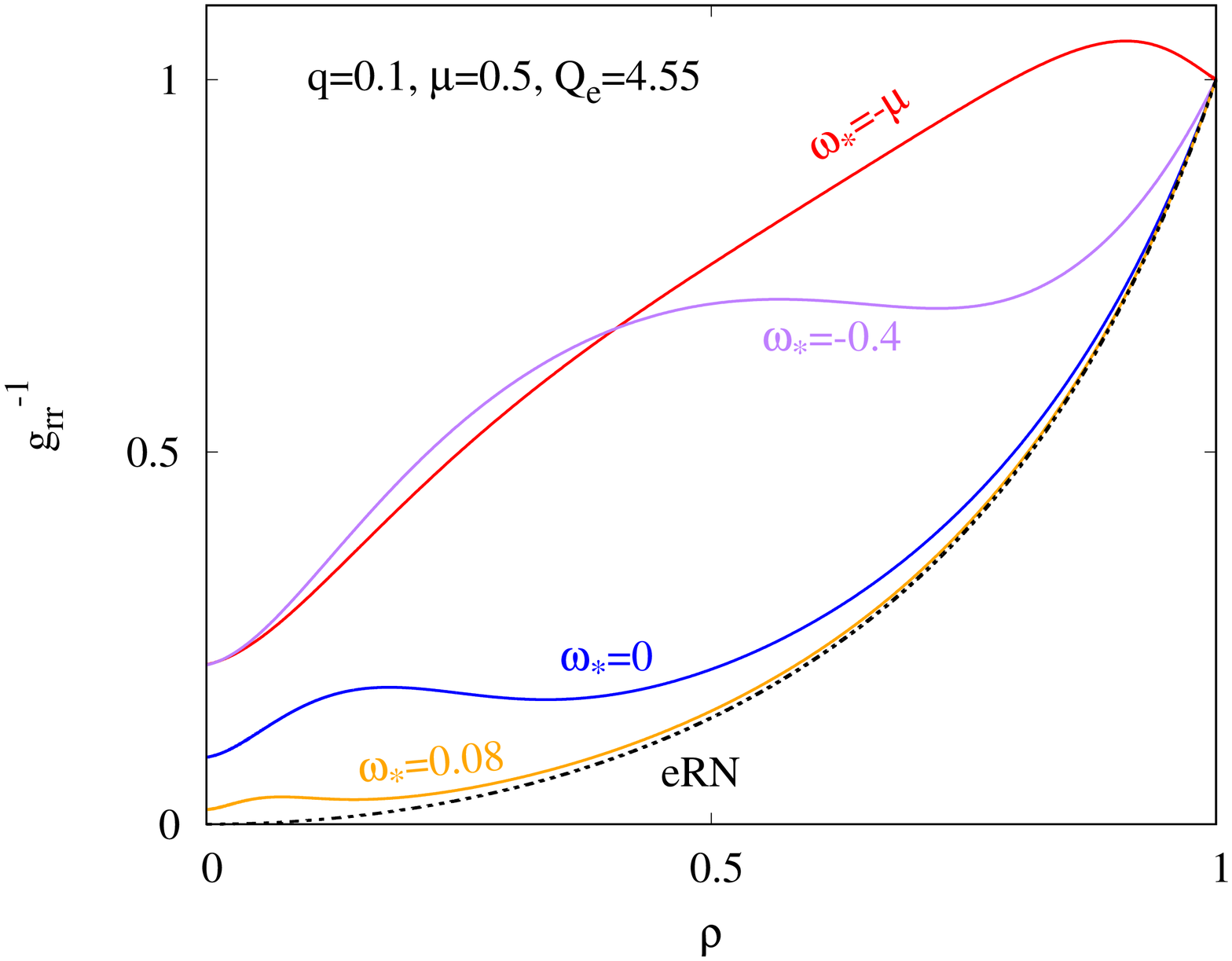}  
	\caption{  
		Metric functions $g_{tt}$ (left) and $g_{rr}$ (right), versus the compactified coordinate $\rho$.
  Each color corresponds to a solution with $w_*=\left\{-\mu,-0.4,0,0.08\right\}$ (red, purple, blue and orange respectively), 
	The extremal RN function is shown for comparison with a black dashed curve. 
	All solutions have $q=0.1$, $\mu=0.5$, $Q_e=4.55$, $\kappa=1$.
	Also, in all plots we define  
	$w_*=w+q \Phi$.
	}
	\label{profile_1}
\end{figure}
%%%%%%%%%%%%%%%%%%%%%%%%%%%%%%%%%

Let us start by describing the generic features of the metric and matter functions that characterize these WHs.

In Figure \ref{profile_1} we show the typical profile for the metric functions $g_{tt}$ (left) and $g_{rr}$ (right), versus the compactified coordinate $\rho$ (as defined by the eq. (\ref{rho}) in Appendix \ref{det_num}), with
\begin{eqnarray}
r^2+r_0^2=\frac{r_0^2}{(1-\rho^2)^2}.
\end{eqnarray}  
For these solutions we fix the parameters of the theory to $q=0.1$ and $\mu=0.5$. 
Then WHs can be obtained for fixed values of the electric charge $Q_e=4.55$ and $\kappa=1$, 
with different values of the parameter $w_*=w+q \Phi$.
These are $w_*=\left\{-\mu,-0.4,0,0.08\right\}$, 
being  shown in Figure \ref{profile_1} in red, purple, blue and orange respectively. 
As a comparison we also include the corresponding metric functions for the extremal RN (eRN) 
BH (dashed black curves). 
As we can see in the figure, for $w_*=0.08$, the metric functions overlap those of extremal RN, 
differing only close to the throat.  The numerical results suggest that in the limit $w_*\to q$, 
the metric functions tend to become closer and closer to the extremal RN functions\footnote{In the numerics,
this limit is approach as $w\to 0$ and $\Phi\to 1$.}. 
In the limit, the throat develops a degenerate event horizon, and the solution 
coincides with extremal RN.
 No smooth solutions can be found for $w_*>q$.
 Hence we conclude that extremal RN forms one of the boundaries of the domain of existence of the WHs.

The solutions with $w_*=-\mu$ form the other boundary of the configuration space (red curve). We will refer to these WHs as   `limit' solutions, since the domain of existence cannot be extended beyond this value of the frequency. 
We find that this is a generic feature, also valid for other arbitrary values of the parameters: all the smooth and symmetric WH solutions we have obtained exist only for $-\mu\le w_*<q$.
In many cases, like in the example shown in Figure \ref{profile_1}, these limit configurations possess negative masses, and we will see that this depends on the particular values of $q/\mu$.

%%%%%%%%%%%%%%%%%%%%%%%%%%%%%%%%%%%%%%%%%%%%%%%%%%%%%%%%%%%%%%%%%%%%%%%%%
\begin{figure}
	\centering
	\includegraphics[height=3.0in,angle=-90]{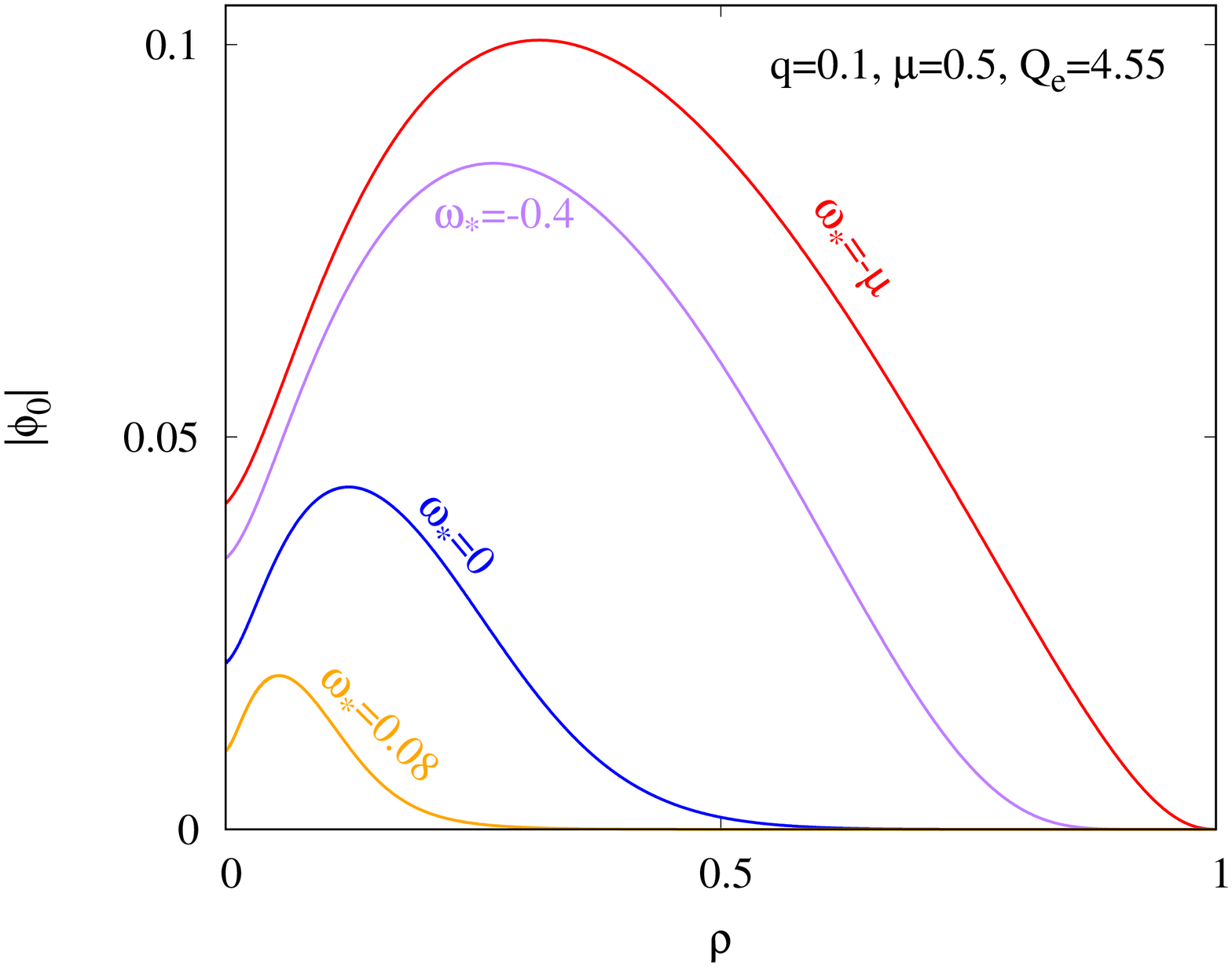} 
	\includegraphics[height=3.0in,angle=-90]{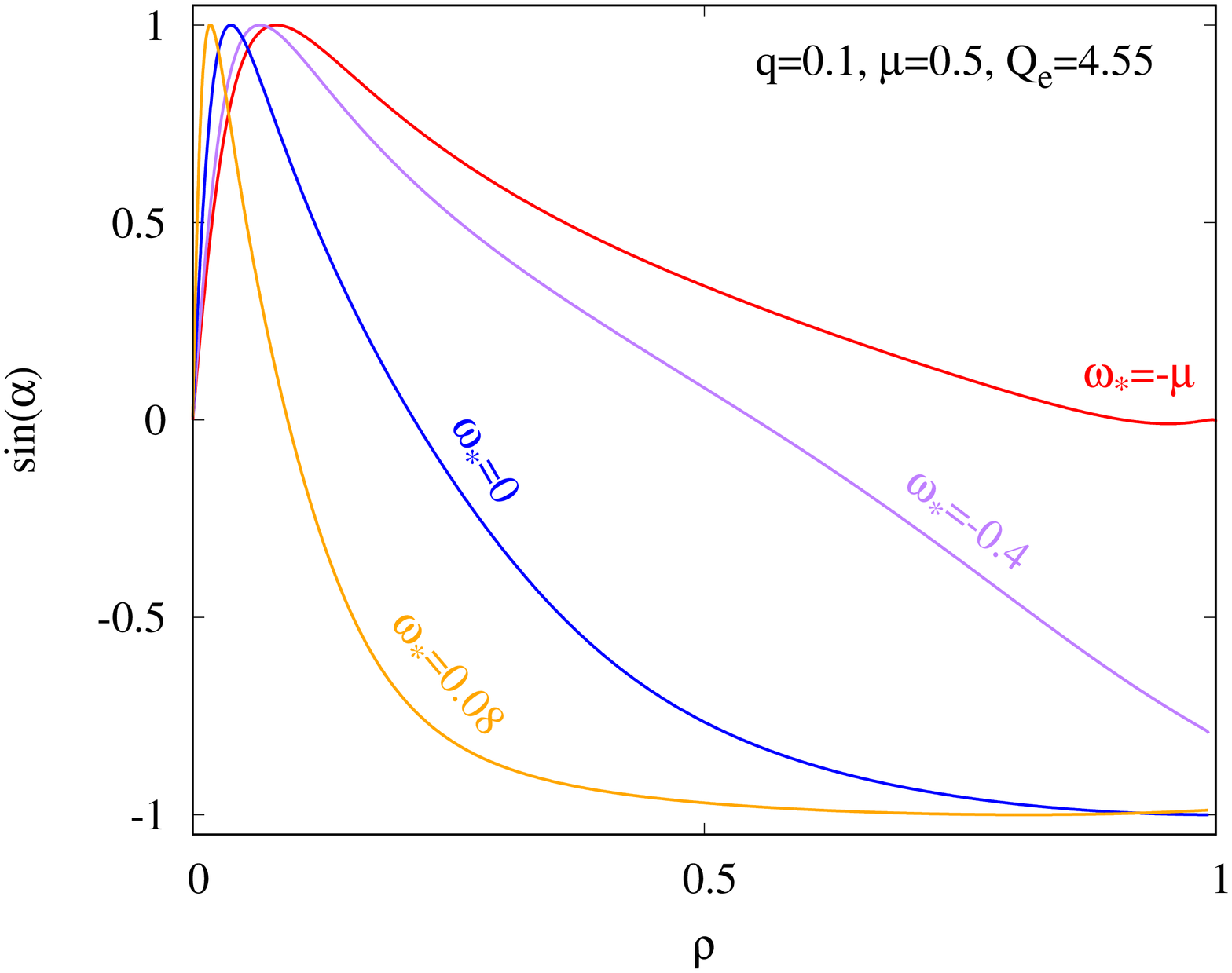}  
	\includegraphics[height=3.0in,angle=-90]{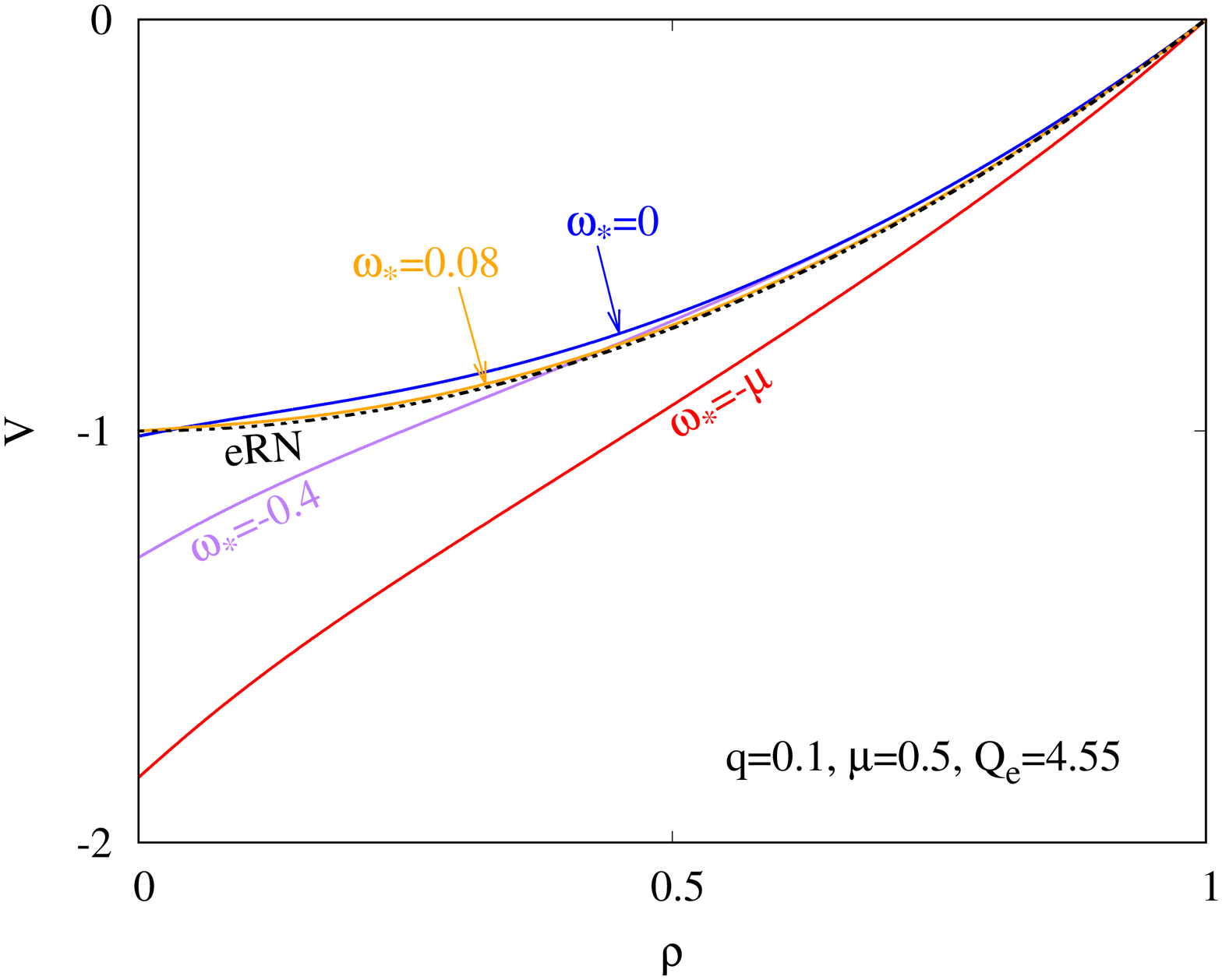}   	
	\includegraphics[height=3.0in,angle=-90]{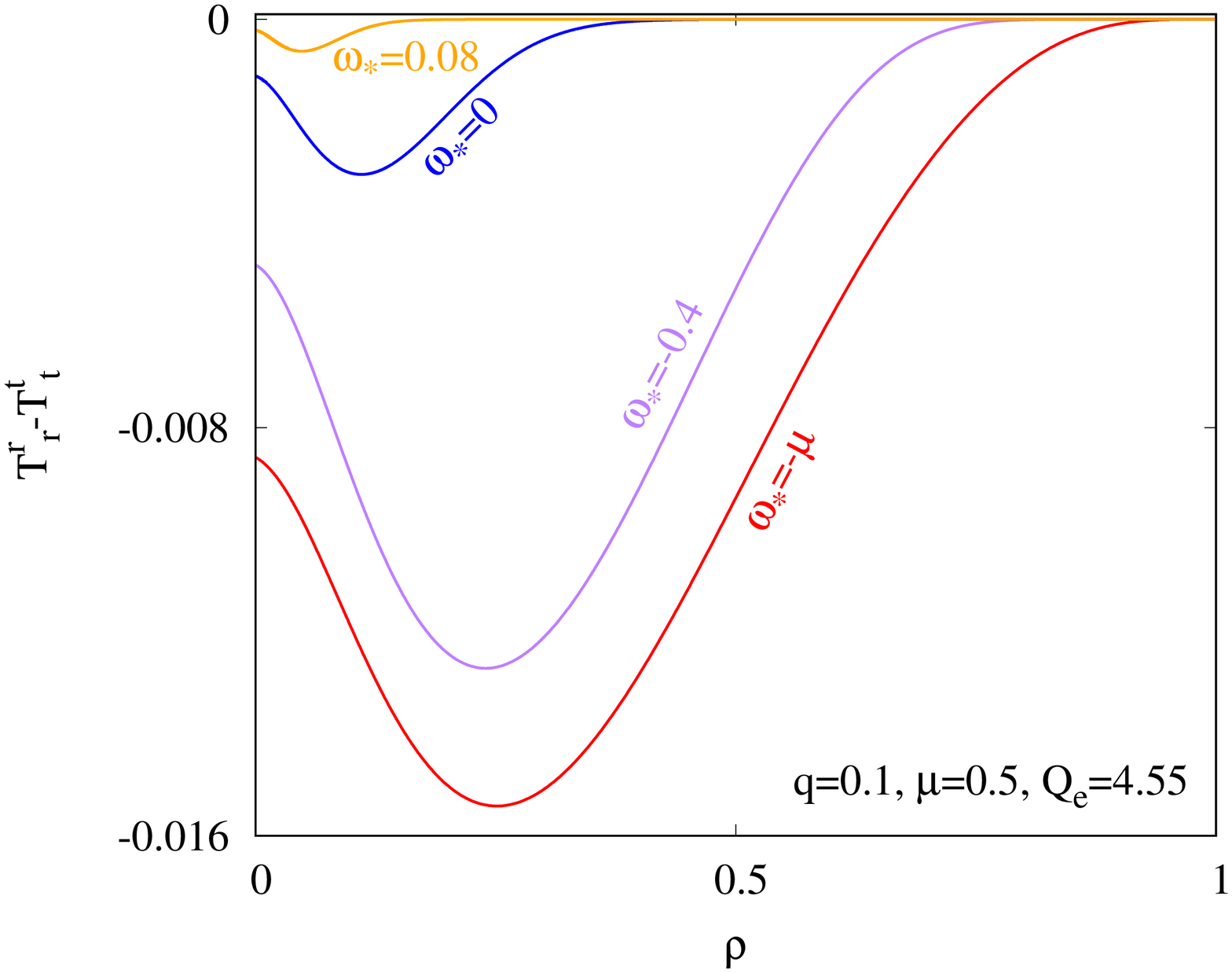}   
	\caption{  
Matter functions $|\phi_0|$ (top left), $\sin{(\alpha)}$ (top right), $V$ (bottom left) and $T_{r}^r-T_{t}^t$ (bottom right) versus the compactified coordinate $\rho$. Each color corresponds to a solution with $w_*=\left\{-\mu,-0.4,0,0.08\right\}$
 (red, purple, blue and orange respectively). All solutions have $q=0.1$, $\mu=0.5$, $Q_e=4.55$, $\kappa=1$.
	}
	\label{profile_2}
\end{figure}
%%%%%%%%%%%%%%%%%%%%%%%%%%%%%%%%%%%%%%%%%%%%%%%%%%%%%%%%%%%%%%%%%%%%%%%%%

Next we discuss the behaviour of the matter content. As an example, in Figure \ref{profile_2} we show the profiles of the matter functions, for the same value of the parameters as in the previous figure. 

The amplitude of the Dirac field $\phi_0$ is shown in Figure \ref{profile_2} (top left).
For all solutions the Dirac field decays exponentially as approaching the asymptotic boundaries. The Dirac field of the 'limit' solutions also decays exponentially at the boundaries (but with a slightly weaker exponential decay, since $\mu_{*}=0$), meaning that these solutions represent also localized states. Hence all the smooth WHs with $-\mu\le w_*<q$ correspond to localized states that can be consistently rescaled to $Q_N=1$ (note that the profiles shown in this section are not rescaled). As $w_*$ is increased towards $w_*=q$, the amplitude of the Dirac field decreases as it shrinks around the throat. 

The other function that characterizes the Dirac spinors is the phase function $\alpha$, which is shown in Figure \ref{profile_2} (top right). The phase varies smoothly as we move away from the throat, and the difference in values between the throat and infinity depends on the value of $w_*$. 

All the smooth configurations we have obtained are necessarily electrically charged, and hence they possess a non-trivial electric potential. We show the electric field function $V$ in Figure \ref{profile_2} (bottom left), where we fix the gauge so that the electric field is zero at $\rho=1$. Again we note that the electric field becomes closer and closer to extremal RN as we increase $w_*\to q$, while it differs the most for the 'limit' configuration.

Another function of interest in order to characterize the matter content of these solutions is $T_{r}^r-T_{t}^t$, obtained from the respective components of the stress-energy tensor. When this function is negative, the null energy condition is violated. We show this function in Figure \ref{profile_2} (bottom right) versus the compactified coordinate $\rho$ where we can see that the null energy condition is violated everywhere. A comparison with the amplitude $\phi_0$ reveals that the violation is maximal close to where the spinor amplitude is larger, which in fact happens slightly outside of the throat.

\begin{figure}
	\centering
	\includegraphics[height=3.0in,angle=-90]{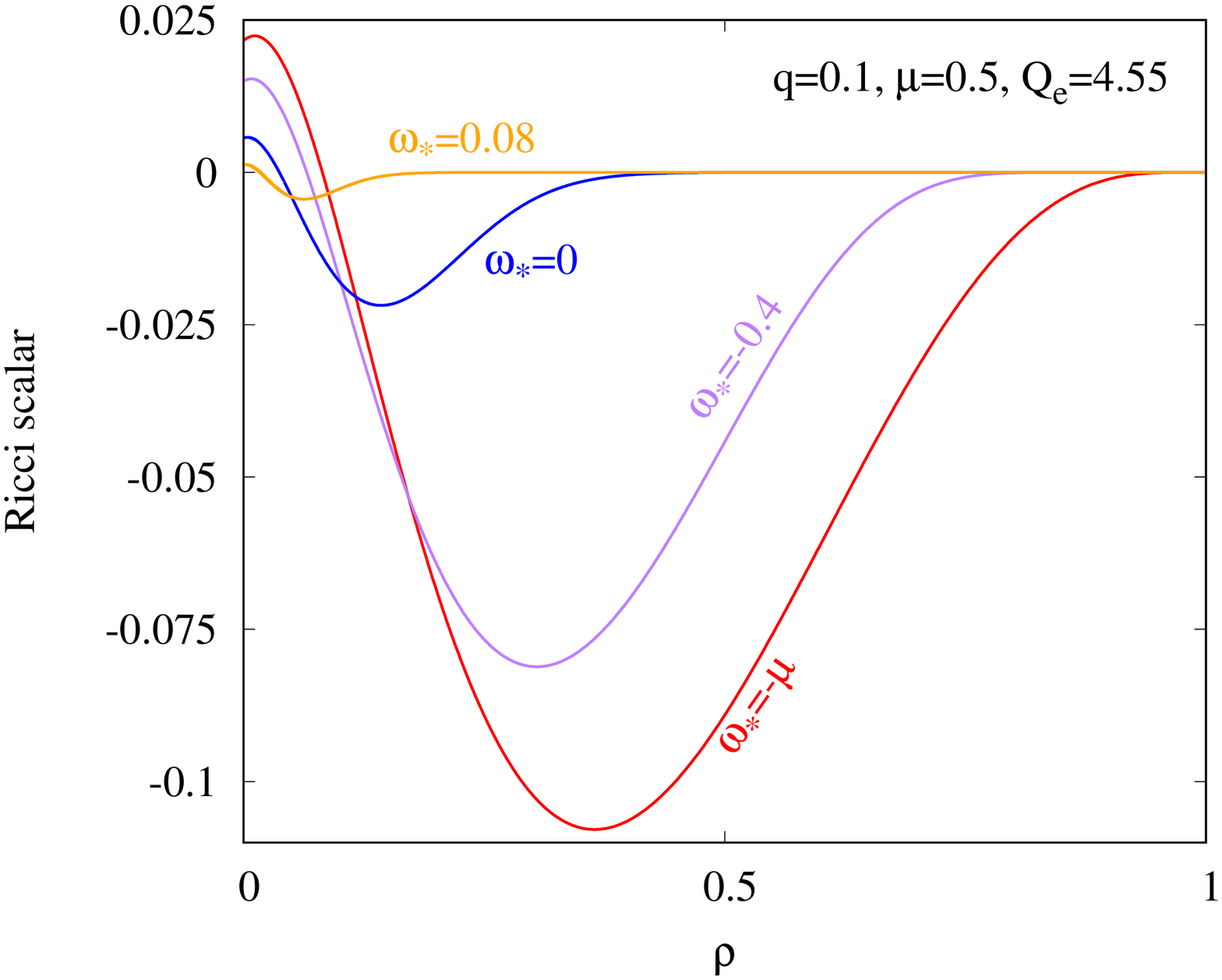}   	
	\includegraphics[height=3.0in,angle=-90]{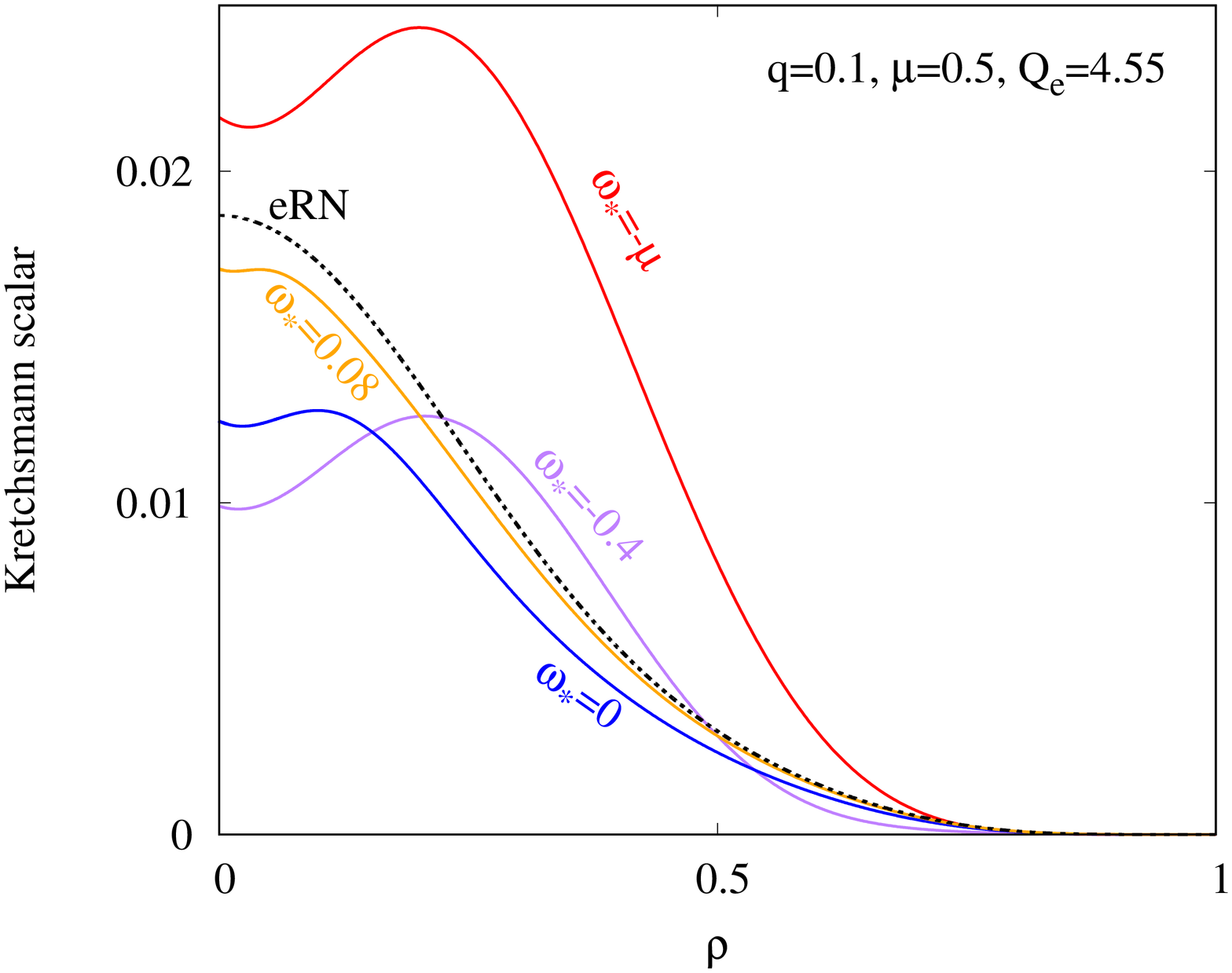}   
	\caption{  
		Ricci scalar (left) and Kretschmann scalar (right) as a function of the compactified coordinate $\rho$. Each color corresponds to a solution with $w_*=\left\{-\mu,-0.4,0,0.08\right\}$ (red, purple, blue and orange respectively). All solutions have $q=0.1$, $\mu=0.5$, $Q_e=4.55$, $\kappa=1$.
	}
	\label{profile_3}
\end{figure}

{  
Let us now consider the curvature invariants for the same configurations we have been discussing. 
In Figure \ref{profile_3} (left) we show the Ricci scalar, and in Figure \ref{profile_3} (right) the Kretschmann scalar. 
One can see that these curvature invariants are everywhere regular for all the configurations with $-\mu\le w_*<q$.
}

%%%%%%%%%%%%%%%%%%%%%%%%%%%%%%%%%%%%%%%%%%%%%%%%%%%%%%%%%%%%%%%%%%%%%%

{

Finally, let us remark that the properties of the profiles described in this Section were
found to be generic for other values of the theory parameters ($q$, $\mu$), electric charge ($Q_e$) and $\kappa$. 
In particular, both the Ricci and Kretschmann scalars are $finite$ and $smooth$ functions everywhere, 
in particular at the throat, $r=0$. 

}

%

%%%%%%%%%%%%%%%%%%%%%%%%%%%%%%%%%%%%%%%%%%%%%%%%%%%%%%%%%%%%%%%%%%%%%%%%%%%%%%
\subsection{Domain of existence and global properties}
%%%%%%%%%%%%%%%%%%%%%%%%%%%%%%%%%%%%%%%%%%%%%%%%%%%%%%%%%%%%%%%%%%%%%%%%%%%%%%

From the previous profiles, we can extract all the global quantities that characterize the WHs: total charge, mass, throat area, $Q_N$, etc. All the solutions we consider can be appropriately rescaled so that $Q_N=1$. 
However,
 for the study of the domain of existence it is convenient to consider appropriate adimensional products of the global quantities.

In the following we explore the properties of solutions with fixed values of the ratio $q/\mu$. It is then possible to generate families of WHs with fixed values of the adimensional electric charge $\mu Q_e$. Solutions with fixed values of $q/\mu$ and $\mu Q_e$ form a 1-parameter family of WHs, that extend from the limit configuration with $w_*/\mu=-1$ to the extremal RN BH with $w_*/q=1$.

\begin{figure}
	\centering
	\includegraphics[height=2.12in,angle=-90]{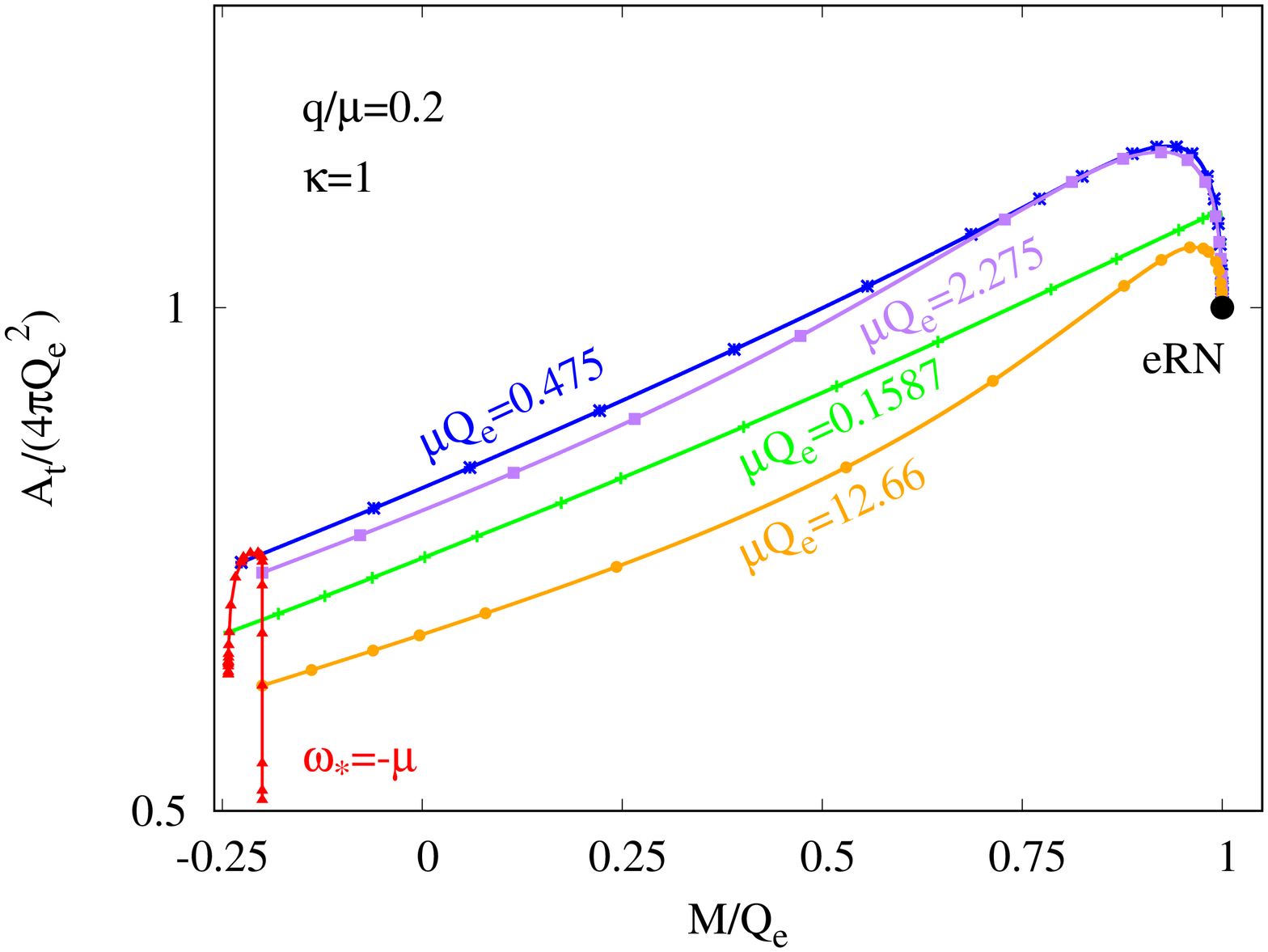}  
	\includegraphics[height=2.12in,angle=-90]{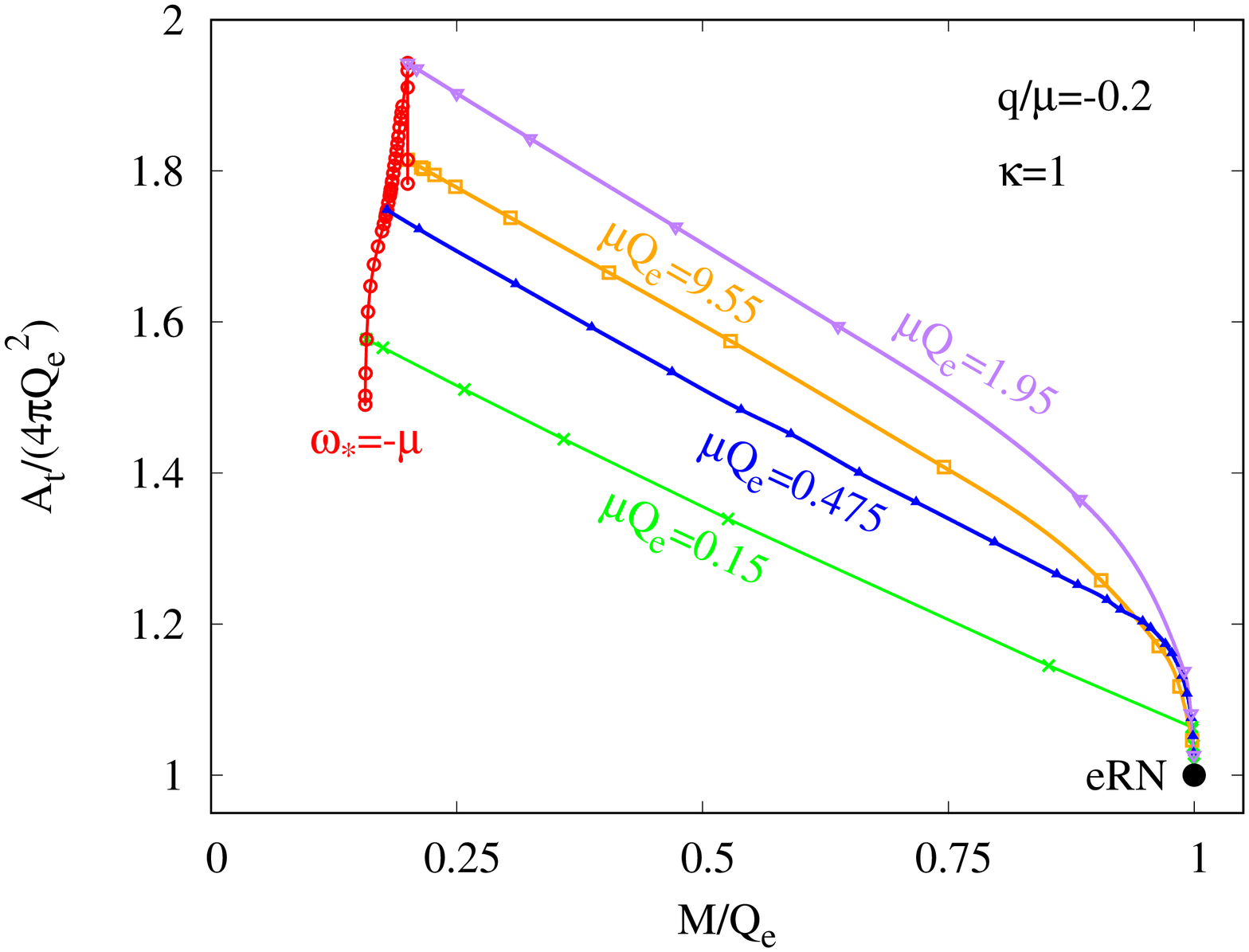}  
	\includegraphics[height=2.12in,angle=-90]{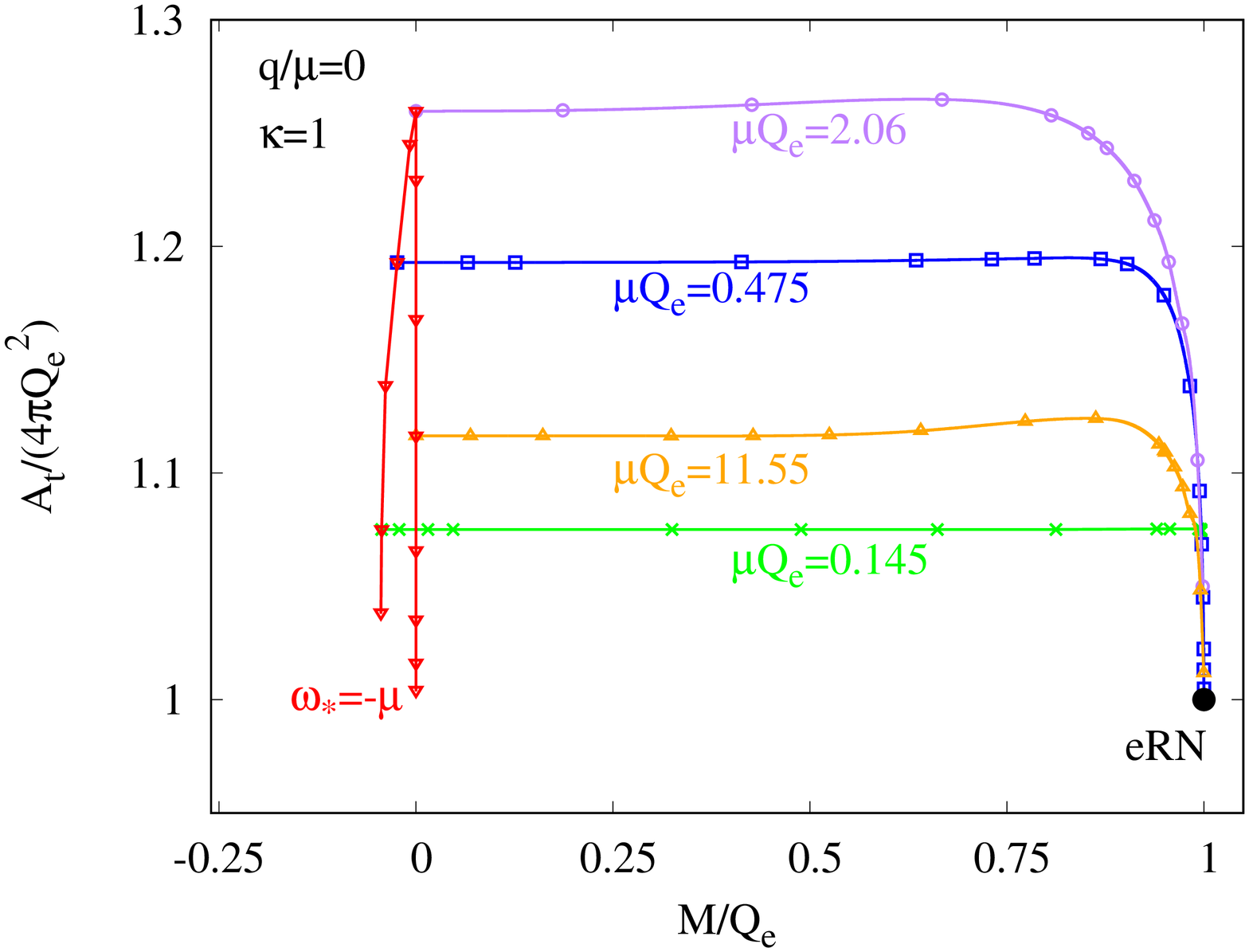} 
	\caption{  
		Throat area scaled to the electric charge, $A_t/(4\pi Q_e^2)$, as a function of the scaled total mass, $M/Q_e$. Top left panel shows solutions for $q/\mu=0.2$, top right panel for $q/\mu=-0.2$, and bottom panel for $q/\mu=0$. Solutions with fixed scaled electric charge $\mu Q_e$ are shown in different colors (orange, purple, blue, green). Limit configurations with $w_*=-\mu$ are shown in red. The black dot represents the extremal RN BH.
		All these WHs are found for  $\kappa=1$.}.
	\label{AtvsQe_1}
\end{figure}

In order to compare the global quantities of these WHs with the extremal RN BH, it is useful to consider global quantities scaled to the electric charge of the configuration. In Figure \ref{AtvsQe_1} we show the scaled throat area, $A_t/(4\pi Q_e^2)$, as a function of the scaled total mass, $M/Q_e$. Top left figure is for $q/\mu=0.2$, top right figure is for $q/\mu=-0.2$, and bottom figure for $q/\mu=0$. Each color curve corresponds to a family of solutions with different values of $\mu Q_e$. The red curve corresponds to the limit configurations with 
$w_*=-\mu$. The black dot indicates the extremal RN BH, for which the scaled horizon area and mass are equal to one. All these solutions have $\kappa=1$.

Branches of solutions with constant $\mu Q_e$ extend in between extremal RN (black dot) and the set of limit solutions (red curve). The ratio $M/Q_e$ is maximal at the extremal RN solution ($M/Q_e=1$), while the smallest mass possible is found for configurations on the limit curve. Depending on the theory, i.e. the value of $q/\mu$, these limit masses can take negative values. Note that for $q/\mu=0.2$ (top right), all the limit configurations have relatively large negative masses. In fact, for large values of $\mu Q_e$, the solutions on the limit curve tend to $M/Q_e=-q/\mu$. The solutions form a vertical line where the area decreases with increasing $\mu Q_e$, while the mass-charge ratio is essentially fixed to $-q/\mu$.

Regarding the area, in Figure \ref{AtvsQe_1} we can see that solutions sufficiently close to extremal RN possess throat areas larger than the corresponding horizon area of the extremal RN BH. In fact this happens for all mass-charge 
ratios in models
 with $q/\mu\le0$. 
Only in models with $q/\mu>0$, it is possible to obtain configurations that possess throat areas smaller than the extremal RN horizon area.

\begin{figure}
	\centering
	\includegraphics[height=3.0in,angle=-90]{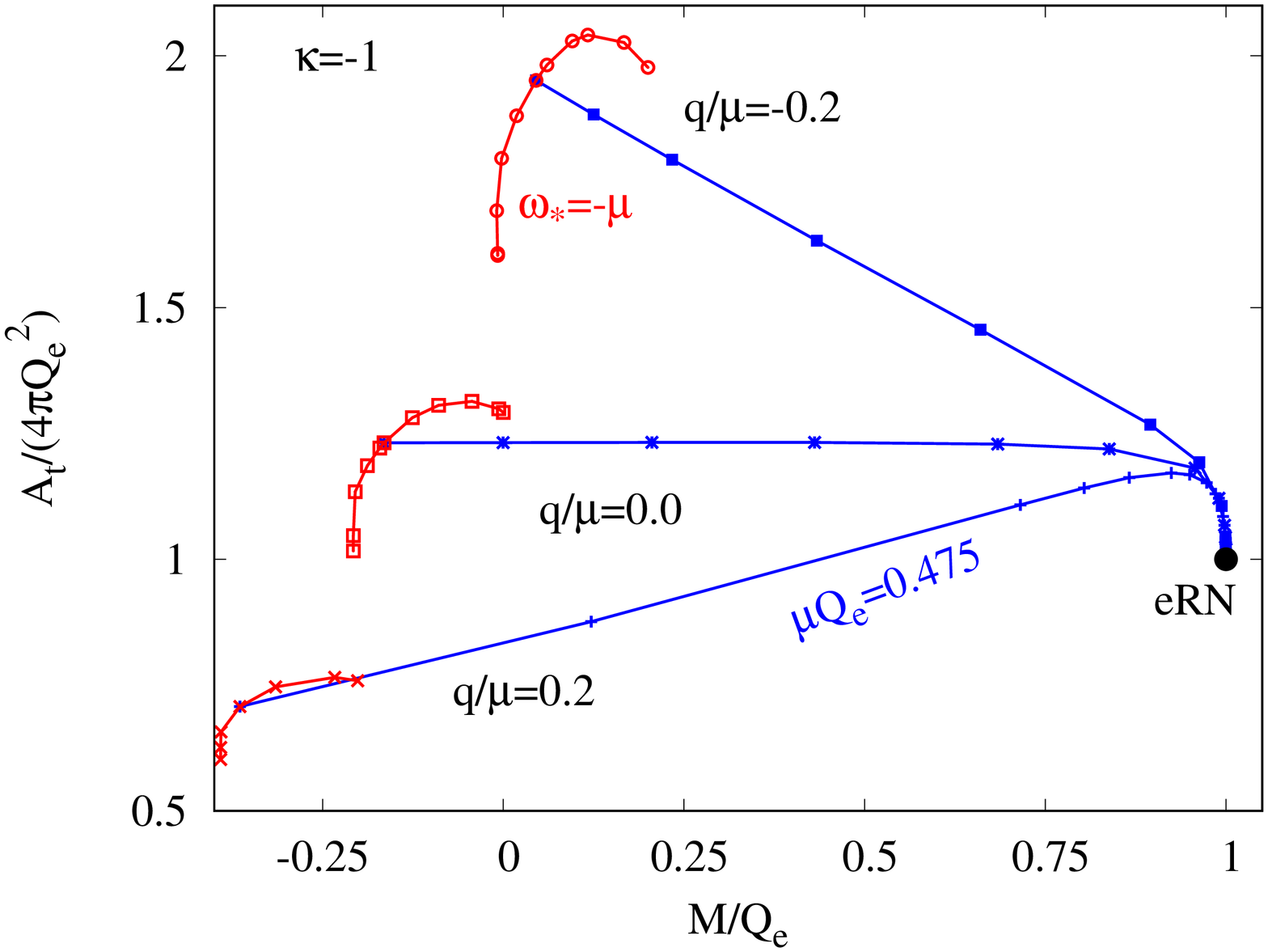} 
	\includegraphics[height=3.0in,angle=-90]{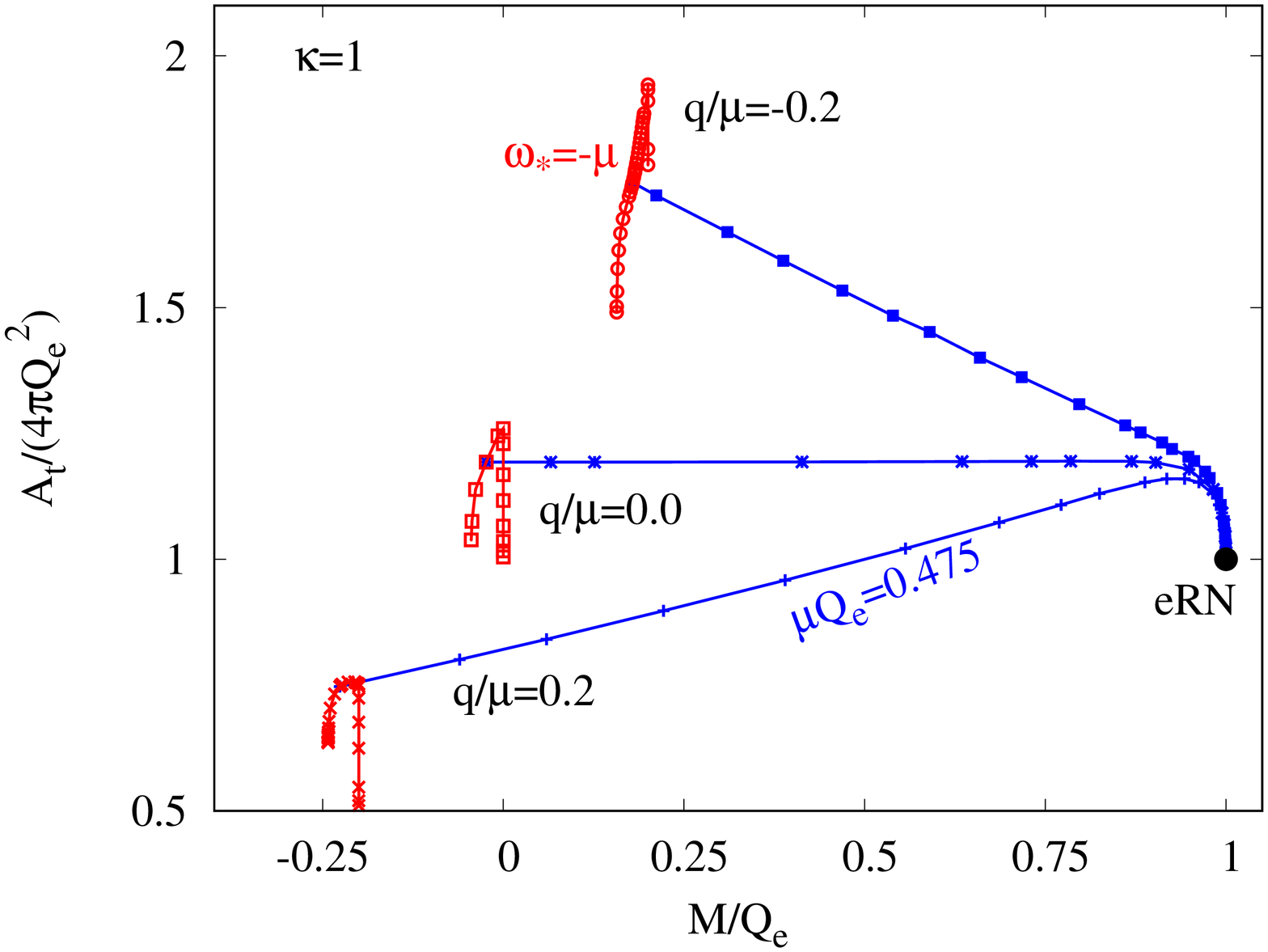}  
	\caption{  
		Throat area scaled to the electric charge, $A_t/(4\pi Q_e^2)$, as a function of the scaled total mass, $M/Q_e$, for $\kappa=-1$ (left) and $\kappa=1$ (right). Limit configurations with $w_*=-\mu$ are shown in red, while the blue curves represent solutions with fixed $\mu Q_e = 0.475$. The black dot indicates the extremal RN BH.
		% All these WHs are found for  $\kappa=1$
		}.
	\label{AtvsQe_2}
\end{figure}

While in Figure \ref{AtvsQe_1} we have fixed $\kappa=1$, the properties of solutions with $\kappa=-1$ do not differ significantly. In Figure \ref{AtvsQe_2} we show again the scaled throat area as a function of the scaled mass. In Figure \ref{AtvsQe_2}(left) we show some subsets of solutions with $\kappa=-1$, and in Figure \ref{AtvsQe_2}(right) we show similar subsets with $\kappa=1$. Qualitatively, the domain of existence is very similar for both values of $\kappa$, the most important differences appearing only close to the limit configurations.

Finally, let us comment that we have not found regular symmetric solutions for 
models with $|q/\mu|>1$, nor WHs with $|M/Q_e|>1$. Such configurations may exist in a more general setting, for instance when considering asymmetric WHs.

%%%%%%%%%%%%%%%%%%%%%%%%%%%%%%%%%%%%%%%%%%%%%%%%%%%%
\subsection{Isocharge ensembles}
%%%%%%%%%%%%%%%%%%%%%%%%%%%%%%%%%%%%%%%%%%%%%%%%%%%%
The previous analysis in terms of adimensional quantities is useful in order to understand the domain of existence of these WHs. It also allows us to compare the properties of the WHs relative to the ones of the extremal RN BH, which plays a key role as it is one boundary of the space of solutions. 

In order to make contact with previous analysis of Finster-Smoller-Yau solitons, in the following we will discuss the properties of the WHs in terms of the Plank scale. To do so we rescale all quantities to $Q_N=1$, and look at ensembles of solutions with fixed values of the electric charge $Q_e$. These ensembles form again a 1-parameter family of solutions, characterized by the spinor mass $\mu$.

\begin{figure}
	\centering
	\includegraphics[height=3.0in,angle=-90]{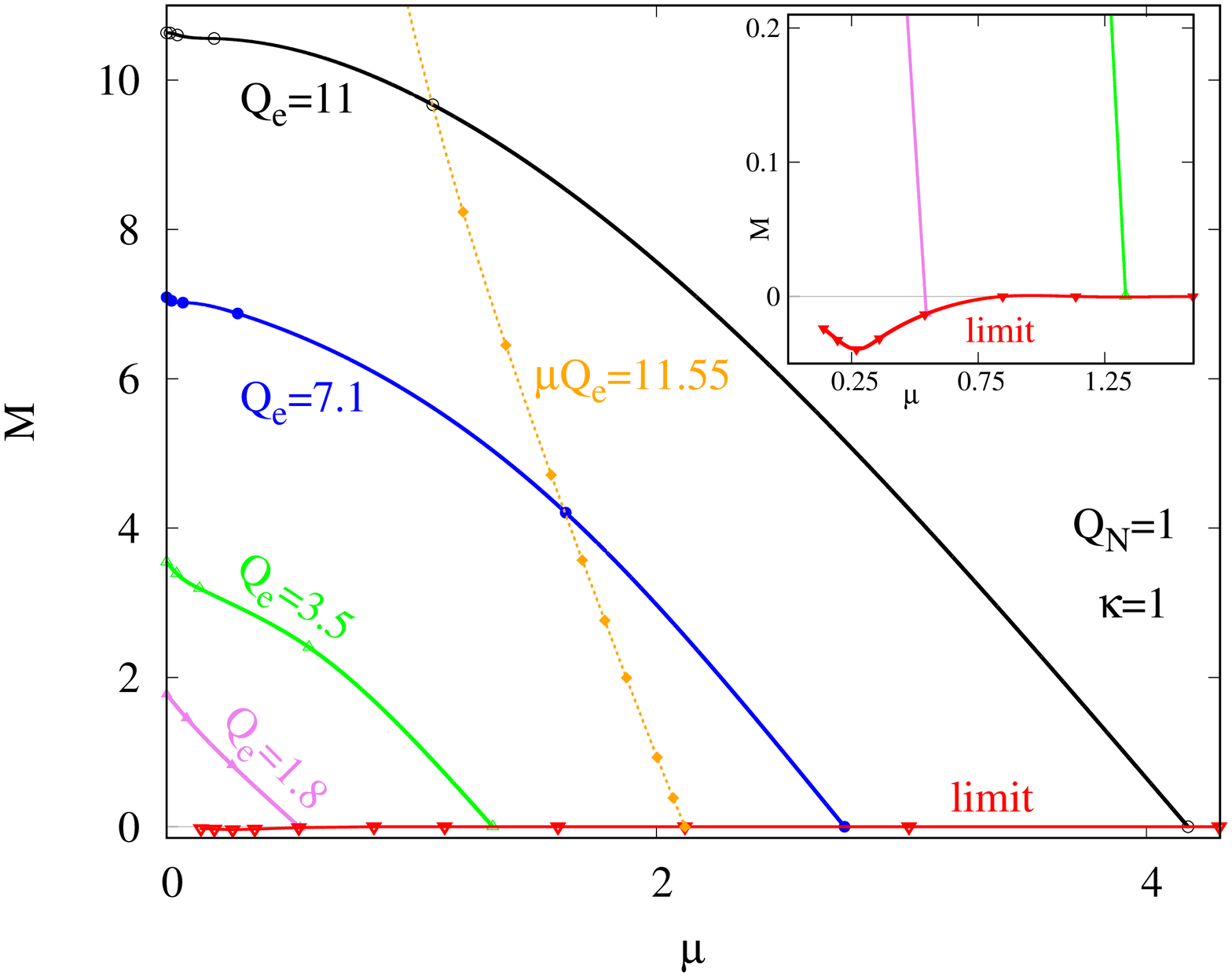} 
	\includegraphics[height=3.0in,angle=-90]{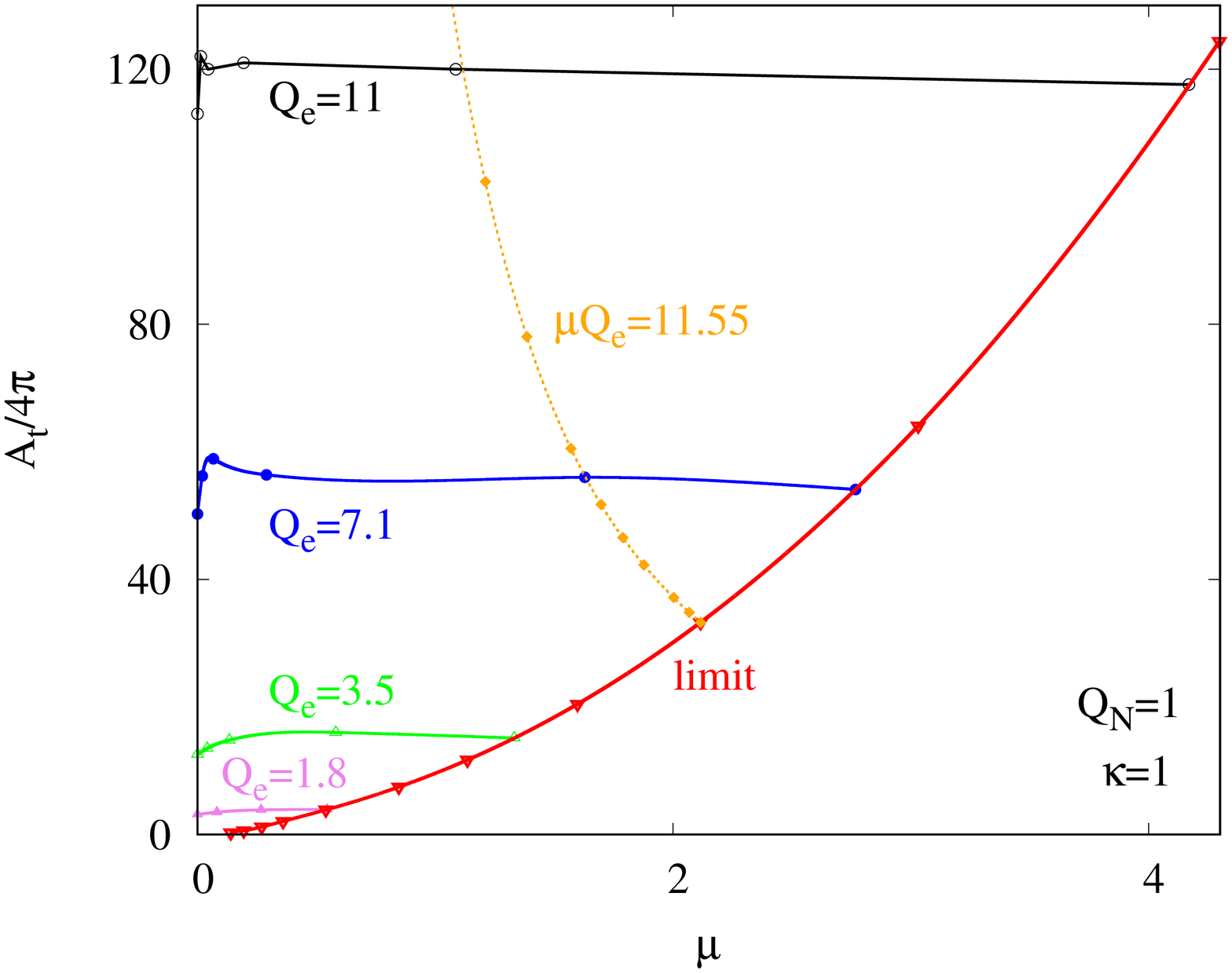}  
	\caption{  
		(left) Mass $M$ vs $\mu$ in Plank units  for solutions with $q=0$. The color curves correspond to families of solutions with fixed electric charge ($Q_e=1.8,3.5,7.1,11$ in pink, green, blue and black respectively). 
		In red we show the limit configurations with $w_*=-\mu$. In orange we include the solutions with fixed $\mu Q_e=11.55$ for comparison with Figure \ref{AtvsQe_1}. (right) Same figure for the throat area $A_t$ as a function of $\mu$. 
	}
	\label{QN1_ungauged}
\end{figure}

In Figure \ref{QN1_ungauged}(left) we show the mass of the WH as a function of the mass of the fermion. For simplicity here we focus on the ungauged case with $q=0$. The solid color curves represent the ensembles of fixed electric charge ($Q_e=1.8,3.5,7.1,11$ in pink, green, blue and black respectively). The red curve corresponds to the limit set, for which $w_*=-\mu$. Along this curve, the electric charge increases with the fermion mass. For reference, we also include the dotted orange curve, representing the family of configurations with fixed $\mu Q_e=11.55$ that is shown in Figure \ref{AtvsQe_1}(bottom).

The WH solutions bifurcate from $\mu=0$, for which we have seen that there is no fermion content, the configuration corresponding to the extremal RN BH with $M=Q$. Along the isocharge ensembles, the mass of the WH decreases monotonically with increasing fermion mass. 
Eventually, a limit value of $\mu$ is reached, for which $w_*=-\mu$ (limit red curve). These solutions possess negative values of the mass, as shown in the inset figure.

Another quantity of interest is the throat area. In Figure \ref{QN1_ungauged}(right) we show the area as a function of the fermion mass in Plank units, for the same sets of solutions as for Figure \ref{QN1_ungauged}(left). We can see that for the isocharge ensembles, the area of the WH throat does not deviate considerably from the horizon area of the extremal RN BH (again at $\mu=0$).

\begin{figure}
	\centering
	\includegraphics[height=3.0in,angle=-90]{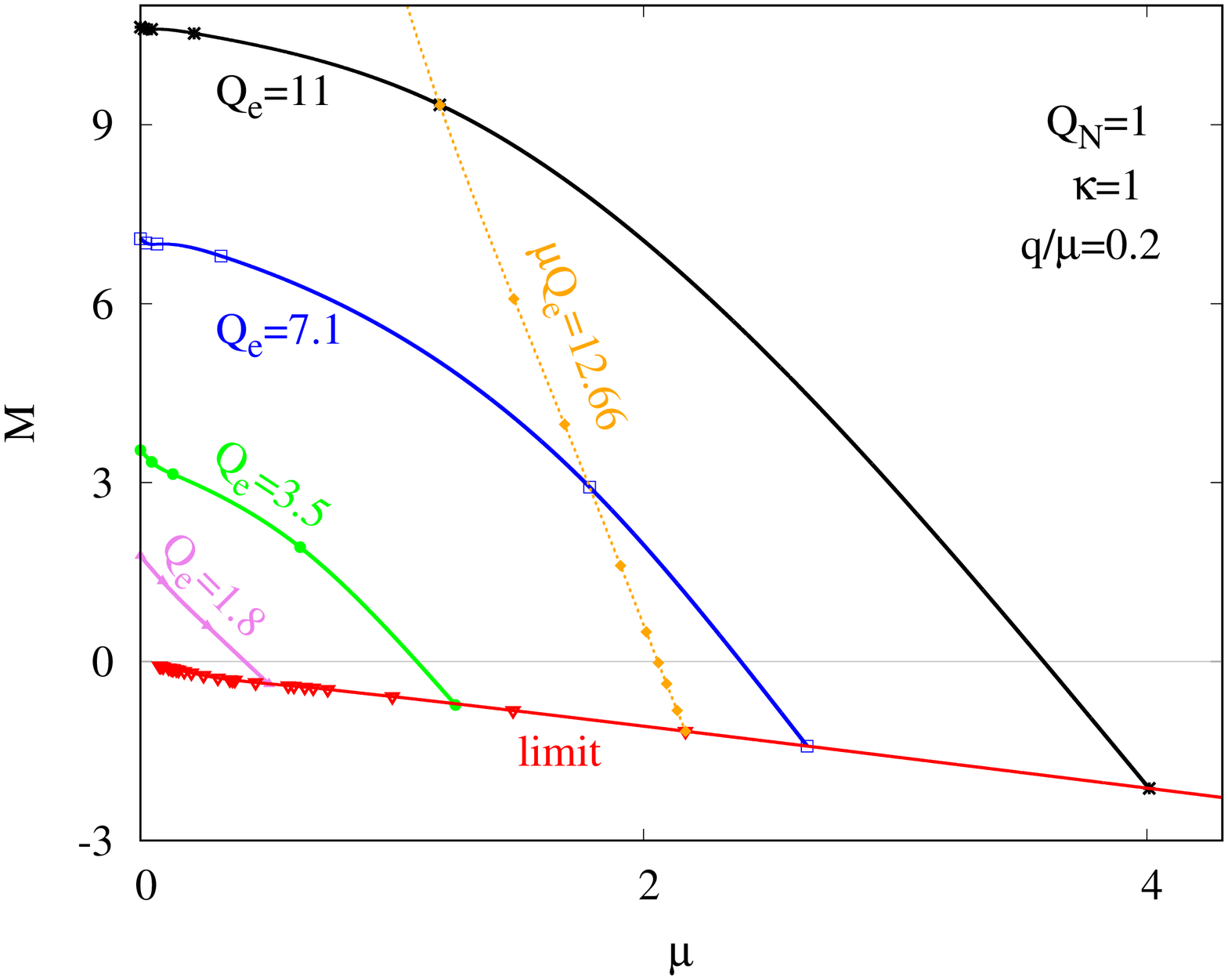} 
	\includegraphics[height=3.0in,angle=-90]{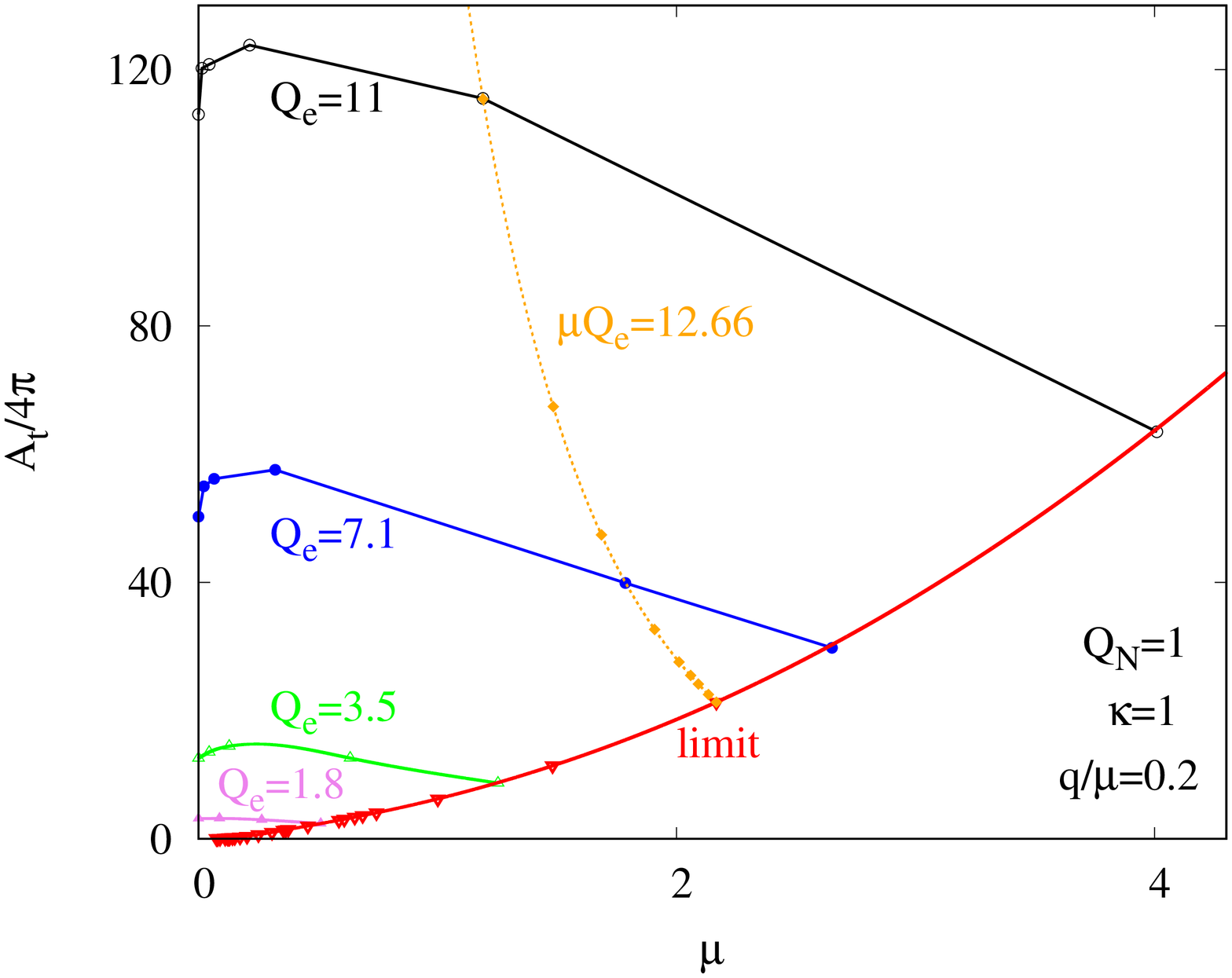}  
	\caption{  
		(left) Mass $M$ vs $\mu$ in Plank units for solutions with $q/\mu=0.2$. The color curves correspond to families of solutions with fixed electric charge ($Q_e=1.8,3.5,7.1,11$ in pink, green, blue and black respectively).
		In red we show the limit configurations with $w_*=-\mu$. In orange we include the solutions with fixed $\mu Q_e=12.66$ for comparison with Figure \ref{AtvsQe_1}. (right) Same figure for the throat area $A_t$ as a function of $\mu$. 
	}
	\label{QN1_gauged}
\end{figure}

For gauged solutions, the behaviour is qualitatively very similar. We show this in Figure \ref{QN1_gauged}, that corresponds 
to models with $q/\mu=0.2$. The main difference occurs close to the limit configurations (the red curve), for which the mass take relatively large negative values, as compared with the ungauged case. On the other hand, for sufficiently large values of the fermion mass, the throat area is slightly smaller than the horizon area of the extremal RN BH.

Figures \ref{QN1_ungauged} and \ref{QN1_gauged} indicate that, for a fixed value of the fermion mass, there is a minimum charge below which it is not possible to form a smooth symmetric WH. For this minimum charge, the mass is also minimal, while the value of the throat area is always of the order of magnitude of the extremal RN BH with the corresponding electric charge.

The results also indicate that it is possible to have WH solutions with arbitrarily large mass and charge, and relatively small values of the fermion mass. The geometry of such WHs do not differ much from the geometry of extremal RN, with the main differences occurring only close to the throat. 

%

 %%%%%%%%%%%%%%%%%%%%%%%%%%%%%%%%%%%%%%%%%%%%%%%%%%%%%%%%%%%%%%%%%%%%%%%%%%%%%%
\section{Further remarks}
%%%%%%%%%%%%%%%%%%%%%%%%%%%%%%%%%%%%%%%%%%%%%%%%%%%%%%%%%%%%%%%%%%%%%%%%%%%%%%

The main purpose of this work was to provide a detailed description of the construction and of the (basic)
 properties of a new type of WH solutions reported in Ref. 
\cite{Blazquez-Salcedo:2020czn}.
As with their 
Finster-Smoller-Yau
solitonic counterparts  
\cite{Finster:1998ws},
\cite{Finster:1998ux},
these Einstein-Dirac-Maxwell (EDM) configurations
are spherically symmetric, with
two massive fermions in
a singlet spinor state.
Also, they are free of singularities, representing localized states.
with a finite mass $M$ and electric charge $Q_e$.

One should remark that the existence of two asymptotic regions for a  WH geometry  
introduces a number of complications in the formulation
of a consistent ansatz, as compared to the
solitonic case.
The main difficulties can be traced back to the fact that,
different from the case of bosonic fields,
a Dirac field necessarily implies a tetrad choice,
with the existence of some special features at the WH throat \cite{Cariglia:2018rhw}.
Another complication originates in choosing to study 
 WH geometries which are symmetric with respect to a reflection at the throat.

Interestingly, all symmetric solutions are overcharged, 
their systematic  analysis revealing that the extremal  Reissner-Nordstr\"om (RN) 
BHs play  an important role,
 as providing 
one of the boundaries of the domain of existence, 
for which the mass/charge ratio is maximal. 
The other boundary of the domain of existence is given by a set of limit configurations, 
for which the mass/charge ratio is minimal. 
We have shown that, in principle, one can obtain WH solutions with bulk geometries very similar to extremal RN, 
only significantly different close to the throat, which is supported by the fermionic matter. 
On the other hand, we find that the throat area of the WHs 
we have considered do not differ significantly 
with respect the horizon area of a extremal RN BH of the same charge. 

\medskip

Let us close this Section with a discussion of 
possible issues and open questions on the subject of WHs in
EDM theory.
First, 
a better understanding of the 
behaviour 
at the 
WH's
throat
of the metric and matter functions 
is 
clearly
necessary.
For example,
as mentioned in Section 3, 
the first derivative
of the radial metric function is discontinuous at the WH throat, 
for all solutions in this work.
However, this is likely a
consequence of the assumption of reflection symmetry, and 
one expect this feature to be absent for {\it asymmetric} WH geometries.
 
On the other hand, it is well known that WHs generically possess 
dynamical instabilities, 
already at the level of spherically symmetric perturbations
 \cite{Shinkai:2002gv,Gonzalez:2008wd,Gonzalez:2008xk,Cremona:2018wkj,Blazquez-Salcedo:2018ipc}.
Therefore it is possible that the EDM WH also possess similar instabilities.  

Finally, the most challenging issue is to understand the physical relevance of 
this type of solutions.
As with the Finster-Smoller-Yau solitons
\cite{Finster:1998ws},
\cite{Finster:1998ux},
the construction here employs a
 semiclassical approach.
That is, the Dirac-Maxwell and 
Einstein equations are coupled,
the fermionic nature of the Dirac field   
being imposed at the level of the occupation number, only.
The debate on the physical validity of this approach 
has a long history
(see $e.g.$ the  discussion  between Wheeler  and de Witt
in Ref.  \cite{book}, p. 143).
While 
a final answer here is absent in the literature,
one expects that 
the inclusion of quantum corrections to the Dirac stress-energy tensor 
\cite{Parker:2009uva,Groves:2002mh}
 may affect the properties of the solutions 
or even invalidate them 
(if they are on the same order 
of magnitude (or larger)
as those found within 
the  {\it quantum wave function} approach).
 However, there are also arguments that the 
employed  treatment 
(without a second quantization of the Dirac field)
may provide a reasonable
 approximation under certain conditions, 
see $e.g.$ Refs. 
\cite{ArmendarizPicon:2003qk},
\cite{Finster:1999xc}.
Moreover,
we expect EDM WHs to exist as well
 in a more complete setting, with a fully quantized (gauged) Dirac field,
as suggested by the results in \cite{Maldacena:2018gjk}.

%%%%%%%%%%%%%%%%%%%%%%%%%%%  
\section*{Acknowledgements}
%%%%%%%%%%%%%%%%%%%%%%%%%%%
%
We would like to thank D. Danielson, G. Satishchandran, R. Wald and R. Weinbaum for insightful remarks on a first version of this draft.
The work of E. R. is supported by the Fundacao para a Ci\^encia e a Tecnologia (FCT) 
project UID/MAT/04106/2019 (CIDMA) and by national funds (OE), through FCT, I.P., in the scope of the framework contract foreseen in the numbers 4, 5 and 6
of the article 23, of the Decree-Law 57/2016, of August 29,
changed by Law 57/2017, of July 19. We acknowledge support  from the project PTDC/FIS-OUT/28407/2017 and PTDC/FIS-AST/3041/2020.  
 This work has further been supported by  the  European  Union's  Horizon  2020  research  and  innovation  (RISE) programmes H2020-MSCA-RISE-2015
Grant No.~StronGrHEP-690904 and H2020-MSCA-RISE-2017 Grant No.~FunFiCO-777740. 
The authors would like to acknowledge networking support by the
COST Actions CA15117 {\sl CANTATA} 
and CA16104 {\sl GWverse}. JLBS gratefully acknowledges support by the
DFG Research Training Group 1620 {\sl Models of Gravity} and the DFG project BL 1553.

\appendix

%%%%%%%%%%%%%%%%%%%%%%%%%%%%%%%%%%%%%%%%%%%%%%%%%%
\section{Dirac field: the general formalism  }
\label{Dfld_frm}
\setcounter{equation}{0}
\renewcommand{\theequation}{A.\arabic{equation}}
%%%%%%%%%%%%%%%%%%%%%%%%%%%%%%%%%%%%%%%%%%%%%%%%%%

%%%%%%%%%%%%%%%%%%%%%%%%%%%%%%%%%%%%%%%%%%%%%%%%%%
%\subsection{Conventions}
%%%%%%%%%%%%%%%%%%%%%%%%%%%%%%%%%%%%%%%%%%%%%%%%%%

In what follows we shall use the 
 conventions and notation used  in Ref. \cite{Dolan:2015eua}. 
Coordinate indices are denoted with Greek letters $\alpha, \beta, \gamma \ldots$ 
and tetrad basis indices with Roman letters $a,b,c,\ldots$. 
Also,  $\partial_\mu$, $\nabla_\mu$ and $\hat{D}_\mu$ are used to denote partial, covariant and spinor derivatives, respectively.

One starts by defining a set of four tetrads   $e_a=e_a^\alpha  \frac{\partial} {\partial x^\alpha} $,
with 
\begin{eqnarray}
e_a^\alpha = \{e_0^\alpha, e_1^\alpha, e_2^\alpha, e_3^\alpha \},
\end{eqnarray}
 which we take to be an orthonormal basis, $i.e.$, 
\begin{eqnarray}
g_{\alpha \beta} e_a^\alpha e_b^\beta = \eta_{\alpha \beta},~~~{\rm with}~~\eta_{ab} = {\rm diag}(-1, 1, 1, 1).
\end{eqnarray}
Also
\begin{eqnarray}
e^a_\alpha = \eta^{a b} g_{\alpha \beta} e_b^\beta~~~ {\rm and}~~~g_{\alp\bet} = \eta_{ab} e_{\alp}^a e_{\bet}^b.
\end{eqnarray}  

To define the  gamma matrices, we start by 
introducing 
two sets of $4 \times 4$ matrices $\gamma^{\alpha}$ and $\hat{\gamma}^{a}$
which satisfy the relation
(where $\{A,B\} = A B + B A$):
\begin{eqnarray}
\label{gamanti}
\{ \gamma^\alpha, \gamma^\beta \}  = 2 g^{\alpha \beta} I_4 ,~~~
 \{ \hat{\gamma}^a, \hat{\gamma}^b \}  = 2 \eta^{ab} I_4, 
\end{eqnarray}
with
\beq
\gamma^\alpha =  e_a^\alpha \hat{\gamma}^a ,~~
{\rm and}~~\hat{\gamma}_a = \eta_{ab} \hat{\gamma}^b,~~\gamma_\alpha = g_{\alp \bet} \gamma^\bet.
\eeq
Our choice for th matrices $\tilde{\gam}$ is
\begin{eqnarray}
\tilde{\gam}^0 = \begin{pmatrix} O & I \\ I & O \end{pmatrix}, \quad \tilde{\gam}^i = \begin{pmatrix} O & \sig_i \\ -\sig_i & O \end{pmatrix} , \quad \quad i = 1,2,3,
\end{eqnarray}
where  $\sig_i$ are the Pauli matrices 
\begin{eqnarray}
\sig_1 = \begin{pmatrix} 0 & 1 \\ 1 & 0 \end{pmatrix}, 
\quad 
\sig_2 = \begin{pmatrix} 0 & -\mathrm i \\ \mathrm i & 0 \end{pmatrix}, 
\quad 
\sig_3 = \begin{pmatrix} 1 & 0 \\ 0 & -1 \end{pmatrix} ,
\end{eqnarray}
  $I$ is the $2\times2$ identity and $O$ is the $2 \times 2$ zero matrix.
The matrices $\hat{\gamma}^{a}$
are defined as
\begin{eqnarray}
\hat{\gamma}^1 = \mathrm i \tilde{\gamma}^3,~~
\hat{\gamma}^2 = \mathrm i \tilde{\gamma}^1,~~
\hat{\gamma}^3 = \mathrm i \tilde{\gamma}^2,~~
\hat{\gamma}^0 = \mathrm i \tilde{\gam}^0.
\end{eqnarray}

Furthermore, we  also  define the Dirac conjugate
\begin{eqnarray}
\overline{\Psi} \equiv \Psi^\dagger \alpha,
\end{eqnarray}
where $\alpha = -\hat{\gamma}^0$
and $\Psi^\dagger$ the  Hermitian conjugate of $\Psi$. 
Also, the spinor covariant derivative $\hat{D}_\nu$ is defined as
\begin{eqnarray}
\hat{D}_\nu =  \partial_\nu - \Gamma_\nu ,
\end{eqnarray}
while 
the covariant derivative of the conjugate spinor is 
\begin{eqnarray}
\hat{D}_{\mu} \overline{\Psi} = \partial_\mu \overline{\Psi} + \overline{\Psi} \Gamma_\mu.
\end{eqnarray}
The spinor connection matrices $\Gamma_\nu$ is defined in terms of the spin-connection $w_{\mu a b}$  \cite{Dolan:2015eua}
\begin{eqnarray}
\Gamma_\alpha = -\frac{1}{4} w_{\alpha \, b c} \hat{\gamma}^b \hat{\gamma}^c , ~~{\rm~ with}~~
\tensor{w}{_\mu ^a _b} = e^a_\nu e^\lambda_b \tensor{\Gamma}{^\nu _{\mu \lambda}} - e_b^\lambda \partial_{\mu} e^a_{\lambda} ,
\end{eqnarray}
${\Gamma}{^\nu _{\mu \lambda}} $ being the Christoffel symbols associated with $g_{\alpha \beta}$.
%
%
%
%
%
%
%
%
%
%
%
%
%
%
%
%

 %%%%%%%%%%%%%%%%%%%%%%%%%%%%%%%%%%%%%%%%%%%%%%%%%%
\section{Dirac equation on a spherically symmetric background: separability}
\label{Deq_sphr}
\setcounter{equation}{0}
\renewcommand{\theequation}{B.\arabic{equation}}
 %%%%%%%%%%%%%%%%%%%%%%%%%%%%%%%%%%%%%%%%%%%%%%%%%%

The Dirac operator on the spherically symmetric background (\ref{metric}) takes the form
\begin{eqnarray}
\label{simp_dirac}
\gamma^\nu\hat{D}_\nu
= \frac{\epsilon_t}{F_0} \hat\gamma^t (\partial_t-\mathrm i q V) +   \frac{\epsilon_r}{F_1} \hat\gamma^r \left[ \partial_r + \partial_r (\ln F_2 \sqrt{F_0}) \right] + \frac{\mathrm i}{F_2} \hat\gamma^r \hat\gamma^t \mathcal K \, ,
\end{eqnarray}
with the operator
\begin{eqnarray}
\mathcal K = \mathrm i \hat\gamma^t\hat\gamma^r\left[\hat\gamma^\theta\left(\partial_\theta+\frac{\cos\theta}{\sin\theta}\right)+\frac{1}{\sin\theta}\hat\gamma^\phi\partial_\phi\right]
\end{eqnarray}
being the angular Dirac operator (the Dirac operator on the two sphere \cite{Abrikosov:2002jr}). By construction we have $[\mathcal D, \mathcal K] = 0$ and $[\mathcal D, \partial_t] = 0$.

The Dirac equation (\ref{simp_dirac}) is decoupled when considering the Ansatz (\ref{Dirac-p}, since
\begin{eqnarray}
\mathcal K \Psi^{[1,2]} = \pm \kappa \Psi^{[1,2]} \\
\partial_{t} \Psi^{[1,2]} = -\mathrm i\omega \Psi^{[1,2]}
\end{eqnarray}
Choosing the parametrization (\ref{z}), after some algebraic manipulations, the Dirac equation (\ref{simp_dirac}) can be reduced to the differential equations (\ref{eqD}). We further constrain to spinors with $|\kappa|=1$. Then, by choosing the radial dependence of the spinor $\Psi^{[1,2]}$ as in equation (\ref{z}) further
 simplifies the total stress energy momentum tensor, 
becoming diagonal and compatible with a  spherically symmetric line element \cite{Finster:1998ju,Finster:1998ux,Blazquez-Salcedo:2019uqq}. 
 
%%%%%%%%%%%%%%%%%%%%%%%%%%%%%%%%%%%%%%%%%%%%%%%%%%
\section{Details on the numerical approach}
\label{det_num}
\setcounter{equation}{0}
\renewcommand{\theequation}{C.\arabic{equation}}
%%%%%%%%%%%%%%%%%%%%%%%%%%%%%%%%%%%%%%%%%%%%%%%%%%
   
%%%%%%%%%%%%%%%%%%%%%%%%%%%%%%%%%%%%%%%%%%%%%%%%%%%%%%%%%%%%%%%%%%%%%%%%%%%%%%
%
%%%%%%%%%%%%%%%%%%%%%%%%%%%%%%%%%%%%%%%%%%%%%%%%%%%%%%%%%%%%%%%%%%%%%%%%%%%%%%

In the numerics, we have found convenient to define a new radial (compactified)
coordinate $\rho$, 
\begin{eqnarray}
\label{rho}
 \rho= {\rm sign}(r)\sqrt{1-\frac{r_0}{\sqrt{r^2+r_0^2}}}  
\end{eqnarray}
such that $\rho \to \pm 1$ as $r\to \pm \infty$,
while $\rho=0$ for $r=0$.
Then the line element (\ref{metric}) 
takes the following form, as written in terms of $\rho$
together with a redefinition of the  metric functions,
$F_0=\sqrt{\sigma}$
and 
$F_1=1/\sqrt{n}$:
\begin{eqnarray}
ds^2= -\sigma(\rho) dt^2+  \frac{4r_0^2}{(1-\rho^2)^2}\frac{d\rho^2 }{n(\rho)}
+\frac{r_0^2}{(1-\rho^2)^2}d\Omega^2.
\end{eqnarray}
Also, in numerics we employ two new spinor functions $f,g$, with 
\begin{eqnarray}
P=\frac{1}{\sqrt{2}}(f-g),~~Q=\frac{1}{\sqrt{2}}(f+g).
\end{eqnarray}

With the above redefinitions,
  the Einstein-Dirac-Maxwell eqs. 
	(\ref{eqE}), 
	(\ref{eqD})
	(\ref{eqM})
take the following form
\begin{eqnarray}
&&
\frac{d\sigma}{d\rho}=
-\frac{1-\rho^2}{2\rho}\left(\frac{dV}{d\rho}\right)^2
+ \frac{16 \epsilon_r r_0\sigma}{\sqrt{n}\rho(1-\rho^2)}\left(g\frac{df}{d\rho}-f\frac{dg}{d\rho}\right)
+
\frac{2\sigma(1-n\rho^2)}{n\rho(1-\rho^2)}~,
%)
%
 %
%
\\
&&
\frac{dn}{d\rho}=
\frac{2}{\rho}\frac{1-n}{1-\rho^2}
-\frac{n}{\sigma}\frac{1-\rho^2}{2\rho}\left(\frac{dV}{d\rho}\right)^2
-\frac{32r_0^2\epsilon_t(w+qV)}{\sqrt{\sigma}\rho(1-\rho^2)^3}(f^2+g^2)~,
\end{eqnarray} 

\begin{eqnarray}
\frac{df}{d\rho}=
-\frac{1+3\rho^2n}{2n\rho(1-\rho^2)}f
-\frac{2\epsilon_r\kappa}{\sqrt{n}(1-\rho^2)}f
-\frac{2\epsilon_rr_0\mu}{\sqrt{n}(1-\rho^2)^2}g
+\frac{2r_0\epsilon_r\epsilon_t(w+qV)}{\sqrt{n\sigma}(1-\rho^2)^2}g
+\frac{1-\rho^2}{8\sigma\rho}\left(\frac{dV}{d\rho}\right)^2f \nonumber \\
+\frac{8r_0^2\mu}{n\rho(1-\rho^2)^3}(g^2-f^2)f
+\frac{16\kappa r_0}{n\rho(1-\rho^2)^2}f^2g
-\frac{8r_0^2\epsilon_t(w+qV)}{\sqrt{\sigma}n\rho(1-\rho^2)^3}(f^2+g^2)f~,
\end{eqnarray} 

\begin{eqnarray}
\frac{dg}{d\rho}=
-\frac{1+3\rho^2n}{2n\rho(1-\rho^2)}g
+\frac{2\epsilon_r\kappa}{\sqrt{n}(1-\rho^2)}g
-\frac{2\epsilon_rr_0\mu}{\sqrt{n}(1-\rho^2)^2}f
-\frac{2r_0\epsilon_r\epsilon_t(w+qV)}{\sqrt{n\sigma}(1-\rho^2)^2}f
+\frac{1-\rho^2}{8\rho\sigma}\left(\frac{dV}{d\rho}\right)^2g \nonumber \\
+\frac{8r_0^2\mu}{n\rho(1-\rho^2)^3}(g^2-f^2)g
+\frac{16\kappa r_0}{n\rho(1-\rho^2)^2}g^2f
-\frac{8r_0^2\epsilon_t(w+qV)}{\sqrt{\sigma}n\rho(1-\rho^2)^3}(f^2+g^2)g~,
\end{eqnarray}

\begin{eqnarray}
&&
\frac{d^2V}{d\rho^2}=\frac{32\epsilon_tqr_0^2\sqrt{\sigma}}{n(1-\rho^2)^4}(f^2+g^2) 
\nonumber 
\\
&&
{~~~~~~~~~}
+\left[\frac{1}{\rho}
- \frac{32r_0\kappa}{n\rho(1-\rho^2)^2} fg
+\frac{16r_0^2\mu}{n\rho(1-\rho^2)^3}(f^2-g^2)
+\frac{32r_0^2\epsilon_t(w+qV)}{\sqrt{\sigma}n\rho(1-\rho^2)^3}(f^2+g^2)
\right]\frac{dV}{d\rho}~,
\end{eqnarray} 
 which was used in the numerics.
Let us remark that 
the Einstein eqs. 	(\ref{eqE})
contain also an extra second order equation.
This equation was treated as a constraint, being used to monitor 
the accuracy of the numerical results.

%%%%%%%%%%%%%%%%%%%%%%%%%%%%%%%%%%%%%%%%%%%%%%%%%%%%%%%%%%%%%%%%%%%%%%%%%%%%%%
 %
%%%%%%%%%%%%%%%%%%%%%%%%%%%%%%%%%%%%%%%%%%%%%%%%%%%%%%%%%%%%%%%%%%%%%%%%%%%%%%

The above system of five non-linear coupled differential equations 
for the functions $n,\sigma$ and $f,g,V$
was solved by using the  software package COLSYS   \cite{Ascher:1979iha}.
  This solver employs a collocation method for boundary-value
ordinary differential equations and a damped Newton method of quasi-linearization.
 Typical meshes use around
$10^4$ points in the interval $-1 \leq \rho \leq  1$,
 At each iteration step
a linearized problem is solved by using a spline collocation at Gaussian points.
The typical relative accuracy for the solutions reported here 
was around $10^{-10}$.
 
{In order to solve numerically the previous system of equations,  one has to provide numerical values of a number of input parameters. 
To specify a theory, one should fix the value of $\mu$ and $q$ (fermion mass and charge respectively). 
The vielbein ansatz must be fixed by choosing the signs $\epsilon_t$ and $\epsilon_r$
The spinor depends on the parameters $\kappa$ and $w$, while the throat size is fixed with the value of $r_0$. 
Other parameters are imposed as boundary conditions. For instance, the total electric charge is fixed by requiring $\frac{dV}{d\rho}({1}) = \frac{2Q_e}{r_0}$, and we focus on solutions with $f(0)=0$. In practice, once a seed solution is obtained, we explore the space of solutions by keeping all parameters fixed except $r_0$ and $w$. Deforming the seed solution by changing these two parameters allows us to obtain the WH configurations with $\frac{d\sigma}{d\rho}({0}) = 0$, which are the ones reported in this work.}

 %%%%%%%%%%%%%%%%%%%%%%%%%%%%%%%%%%%%%%%%%%%%%%%%%%
\section{An exact solution}
\label{ex_sol}
\setcounter{equation}{0}
\renewcommand{\theequation}{D.\arabic{equation}}
%%%%%%%%%%%%%%%%%%%%%%%%%%%%%%%%%%%%%%%%%%%%%%%%%%

As remarked in  ref. \cite{Blazquez-Salcedo:2020czn},
 the  
 $q =w=\mu= 0$   
limit of 
the Einstein-Dirac-Maxwell equations 
allows for a simple exact WH solution, which captures some 
basic properties of the more general solutions discussed above.
The expression of this solution has been given in\footnote{The same lime element has been obtain 
in Ref. \cite{Bronnikov:2002rn} as a possible metric in a brane world supported by a
bulk-induced tidal stress-energy tensor. }
  \cite{Blazquez-Salcedo:2020czn}
in Schwarzschild-like coordinates, with $F_2=r$.
For the choice (\ref{gc})
of the metric-gauge,
the functions which enter the line element (\ref{metric})
are
\begin{eqnarray} 
F_0(r)=1-\frac{2Q_e^2r_0}{Q_e^2+r_0^2}\frac{1}{F_2(r)},~~
F_1(r)=\frac{\sqrt{1+\frac{r_0}{F_2(r)}}}{\sqrt{1-\frac{Q_e^2}{r_0F_2(r)}}},~~
F_2(r)=\sqrt{r^2+r_0^2}~,
\end{eqnarray} 
with the matter functions 
\begin{eqnarray} 
P(r)=U_1(r)-U_2(r),~~Q(r)=U_1(r)+U_2(r),~~
V(r)= \pm \frac{2Q_e r_0}{Q_e^2+r_0^2}\sqrt{\big(1-\frac{Q_e^2}{r_0F_2(r)}\big)\big(1-\frac{r_0}{F_2(r)}\big)}~,
\end{eqnarray} 
where
\begin{eqnarray} 
&&
U_1(r)= \frac{c_0} {\sqrt{F_0(r)}}
\left(
{\sqrt{1-\frac{Q_e^2}{r_0F_2(r)}}
-\kappa \sqrt{1-\frac{r_0}{F_2(r)}} } 
\right)^2,~~~
\\
\nonumber
&&
U_2(r)= \frac{\kappa r_0} {32 c_0 (Q_e^2+r_0^2)\sqrt{F_0(r)}}
\left(
{\sqrt{1-\frac{Q_e^2}{r_0F_2(r)}}
+\kappa \sqrt{1-\frac{r_0}{F_2(r)}} } 
\right)^2~.
\end{eqnarray} 
This solution contains three essential parameters $(r_0,~Q_e)$ and $c_0$ (with $r_0>Q_e$),
its mass being
\begin{eqnarray} 
M=\frac{2Q_e^2 r_0}{Q_e^2+r_0^2},
\end{eqnarray} 
(note that $Q_e/M>1$).
One can easily see that the metric and the spinor functions do not change under
the transformation $r\to -r$,
containing even functions of $r$  only, 
while for $V(r)$
one can take 
$V(-r)=-V(r)$
 without any loss of generality (note that $V(0)=0$, while $\Phi=\pm M/Q_e$).
Also, the first derivatives of the metric functions $F_i$
vanish at $r=0$, and thus there is no thin mass
shell structure at the throat.

This
WH geometry is supported by the spinors contribution
to the total energy-momentum tensor, being regular
everywhere (for example, the Ricci scalar vanishes, while the Kretschmann scalar is finite and smooth everywhere).
Also, as $Q_e\to r_0$, the extremal Reissner-Nordstr\"om BH is approached,
the Dirac stress energy tensor vanishing.

However, this solution  possesses
some undesirable features. In particular, 
the spinor functions
 $(P,Q)$ 
do not vanish as $r\to \pm \infty$.
Therefore, the
spinor wave function is not normalizable, 
and one cannot impose the one particle condition, $Q_N=1$.

%%%%%%%%%%%%%%%%%%%%%%%%%%%%%%%%%%%%%%%%%%%%%%%%%%%%%%%%%%%%

%%%%%%%%%%%%%%%%%%%%%%%%%%%%%%%%%%%%%%%%%%%%%%%%%%%%%%%%%%%%
	
\end{document}